\UseRawInputEncoding 
\documentclass[preprint,12pt,fleqn]{elsarticle}

\usepackage{amsmath}
\usepackage{amsthm}
\usepackage{amssymb}
\usepackage{eucal}
\usepackage{mathrsfs}
\usepackage{subfigure}
\usepackage{mathtools} 
\usepackage{nccmath} 
\usepackage{fix-cm}
\usepackage{bm}

\def \Om {\Omega}

\begin{document}

\begin{frontmatter}

\title{Topology optimization of the support structure for heat dissipation in additive manufacturing}
\author[mymainaddress]{Takao Miki\corref{mycorrespondingauthor}}
\cortext[mycorrespondingauthor]{Corresponding author}
\ead{mikit@tri-osaka.jp}
\author[mysecondaryaddress]{Shinji Nishiwaki}
\address[mymainaddress]{Osaka Research Institute of Industrial Science and Technology, 7-1, Ayumino-2, Izumi-city, Osaka, 594-1157, Japan}
\address[mysecondaryaddress]{Department of Mechanical Engineering and Science, Kyoto University C3, Kyotodaigaku-katsura, Nishikyo-ku, Kyoto, 615-8540, Japan}
\begin{abstract}
A support structure is required to successfully create structural parts in the powder bed fusion process for additive manufacturing.
In this study, we present the topology optimization of a support structure that improves the heat dissipation in the building process.
First, we construct a numerical method that obtains the temperature field in the building process, represented by the transient heat conduction phenomenon with the volume heat flux.
Next, we formulate an optimization problem for maximizing heat dissipation and develop an optimization algorithm that incorporates a level-set-based topology optimization.
A sensitivity of the objective function is derived using the adjoint variable method.
Finally, several numerical examples are provided to demonstrate the effectiveness and validity of the proposed method.
\end{abstract}
\end{frontmatter}
\section{Introduction}
Additive manufacturing (AM) is a processing method that creates a 3D object by stacking materials in a layer-by-layer manner from a 3D model\cite{gibson2014additive}.
This method is expected to improve the performance of structural parts because it can create more complicated shapes than conventional manufacturing methods, such as machining, molding, and casting.
In metal AM, laser powder bed fusion (LPBF) is utilized in a wide range of industrial fields, especially in aviation, and its effectiveness has already been observed by an optimized design in terms of performance and cost\cite{emmelmann2011laser}.
However, LPBF is known to cause physical problems, such as residual stress and thermal distortion during the manufacturing process, hindering successful creation of the parts. 
These problems are caused by the thermal effects of laser irradiation, which adversely affect the strength and shape accuracy of the fabricated parts\cite{shiomi2004residual,mercelis2006residual}.
In addition, the region that blocks heat flow, such as overhang, causes not only large thermal distortion, but also microstructure heterogeneity and degraded surface quality \cite{ATZENI2015500,cheng2015deformation,FOX2016131}.
To mitigate this thermal problem, it is necessary to improve the heat dissipation performance such that the heat effect of each layer is reduced by appropriately redesigning the parts or adding support structure.
Furthermore, the support structure needs to be added to the part to avoid increasing the manufacturing time and cost because it will be removed after manufacturing.

Topology optimization \cite{bendsoe1988generating,bendsoe1989optimal} is the most flexible structural optimization method that allows topological changes in addition to shape changes and provides higher performance structures than those obtained by other methods.
Recently, several topology optimization methods that incorporate AM manufacturability have been proposed.
Multi-objective optimization approaches that minimize thermal distortion while minimizing the compliance of the target part exist \cite{wildman2017topology,allaire2018taking,MIKI2021103558}.
These methods mainly focus on thermal distortion without considering heat dissipation.
On the other hand, in the optimizing the support structure, several optimization methods with focus on improving heat dissipation have been proposed.
Allaire et al.\cite{allaire2018optimizing} proposed a stationary heat conduction analysis model in which a constant heat flux was applied to the shape boundary and constructed a level-set-based topology optimization that maximizes heat dissipation.
Wang et al.\cite{wang2020optimizing} proposed a stationary heat conduction analysis model that provides heat flux only to overhanging surfaces and developed an optimization method that maximized heat dissipation on those surfaces. 
The use of these analytical models is computationally inexpensive and suitable for combination with topology optimization.
However, the validity of the analytical models was not rigorously confirmed because the heat source from the laser was applied to each layer in the actual building process, and the temperature distribution for each layer was not considered.
Zhou et al.\cite{zhou2019topology} proposed a transient heat conduction analysis model using a moving heat source representing a laser and constructed an optimization method that minimizes the temperature at selected points.
Although this method can estimate the temperature distribution of each layer, the computational cost is high, and the point to minimize the temperature must be determined in advance.

In recent years, various modelling methods have been developed to predict part-scale residual stress and thermal distortion in LPBF.
These modelling methods can be summarized in three approaches \cite{gouge2019experimental,bayat2020part}: inherent strain, agglomerated laser, and flash heating.
Computational cost and accuracy vary greatly depending on the these methods.
The inherent strain \cite{keller2014new,setien2019empirical,chen2019inherent,liang2019modified,prabhune2020fast} is a linear mechanical analysis in which the strain field is obtained through a calibration experiment or a part-scale thermo-mechanical analysis. Then, the strain field is applied to each build layer.
This method is computationally inexpensive because it does not require nonlinear or coupled analyses.
However, heat flow cannot be considered because there is no thermal information in the mechanical analysis.
In contrast, the agglomerated laser and flash heating, which are based on thermo-mechanical analysis, have a temperature history.
The agglomerated laser \cite{hodge2014implementation,hodge2016experimental,chiumenti2017numerical,ganeriwala2019evaluation,gouge2019experimental} is highly accurate because it takes into account the detailed build process parameters: powder layer thickness, laser spot size, laser power, laser scan speed, and total layer time.
However, because typical runs employ more than 32 CPUs, the computational cost is high, and the model size that can be simulated is relatively small.
The last category, flash heating \cite{zaeh2010investigations,papadakis2014computational,prabhakar2015computational,li2017efficient,li2018scalable,yang2018prediction,zhang2019resolution,bayat2020part}, is a simple modelling method that applies an equivalent volume heat flux to each scaled-up powder layer.
Although scan strategies cannot be considered, residual stress and thermal distortion can be predicted, including the temperature history.
Furthermore, the effectiveness of this method has been experimentally verified.

In this study, we construct a support structure optimization method to maximize the heat dissipation for each build layer based on flash heating.
Specifically, we focus on the thermal analysis part of flash heating and incorporate a process model that can predict the temperature field in the topology optimization.
The remainder of this paper is organized as follows. In Section \ref{sec:2}, we propose a simple analytical model based on the transient heat conduction problem to represent the temperature distribution in the AM building process.
In Section \ref{sec:3}, we incorporate the proposed analytical model into a level-set-based topology optimization to formulate an optimization problem that maximizes the heat dissipation in each layer.
In Section \ref{sec:4}, we construct an optimization algorithm for topology optimization using the finite element method (FEM).
Section \ref{sec:5} presents 2D and 3D design examples to demonstrate the validity and effectiveness of the proposed optimization method.
Lastly, Section \ref{sec:6} concludes the study.
This work is implemented with the open source partial differential equation solver FreeFEM++\cite{MR3043640}.
\section{Analytical model for LPBF process}\label{sec:2}
\subsection{Transient heat conduction under volume heat flux}
\begin{figure}[htbp]
	\begin{center}
		\includegraphics[width=13.0cm]{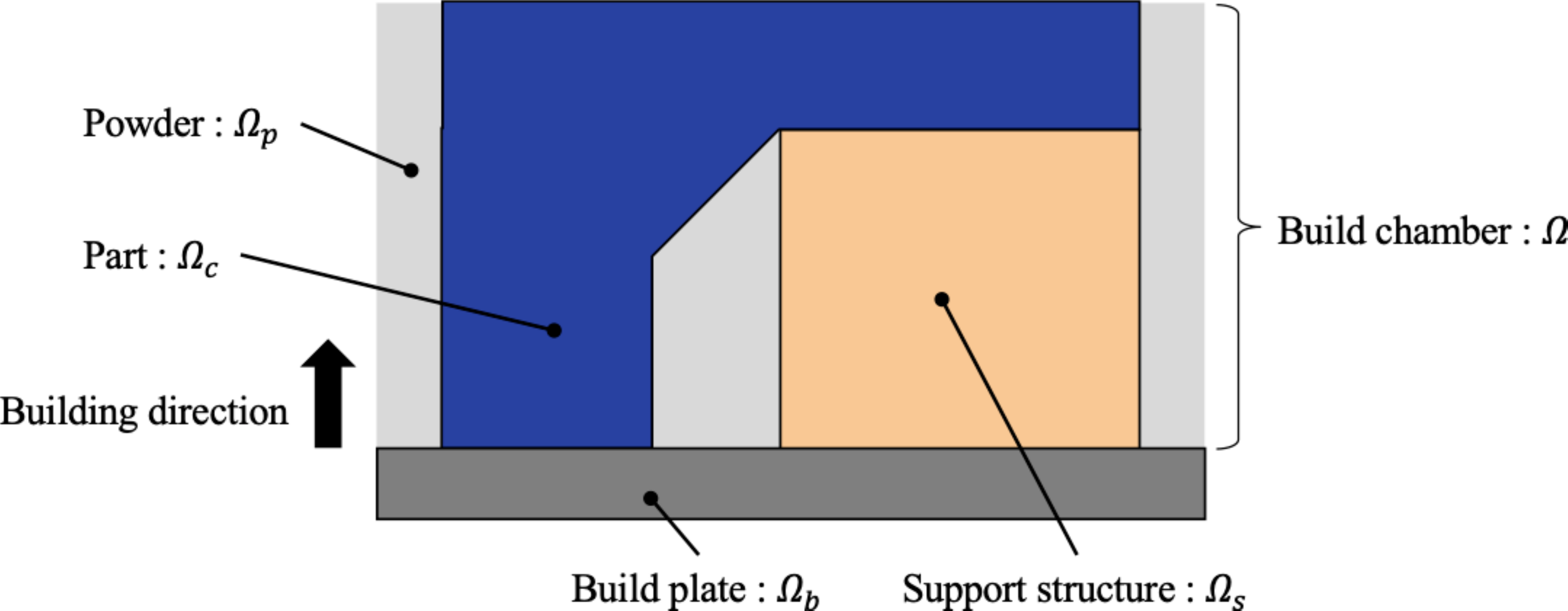}
		\caption{Components of the build chamber in LPBF.}
		\label{fig:buildchamber}
	\end{center}
\end{figure}

\begin{figure}[htbp]
	\begin{center}
		\includegraphics[width=13.0cm]{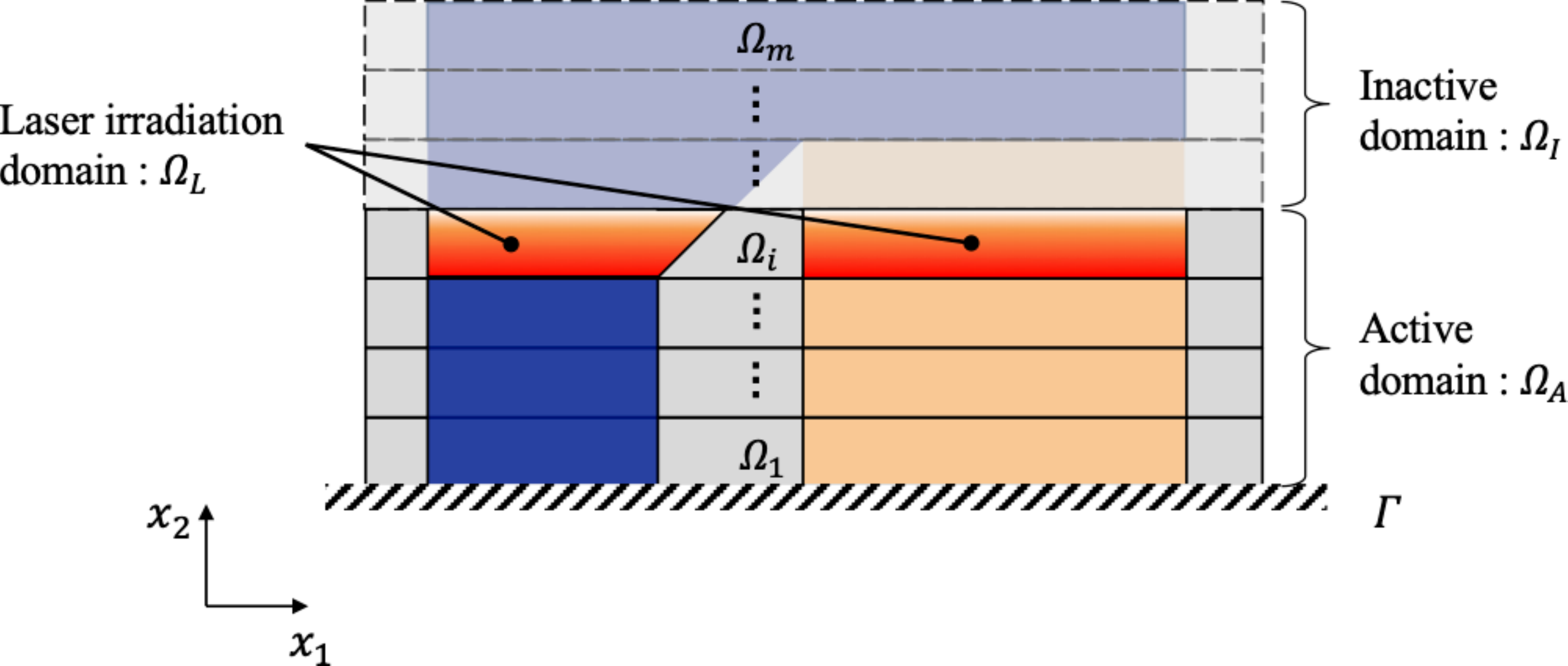}
		\caption{Domains and boundary in the intermediate state of the building process.}
		\label{fig:analysisdomain}
	\end{center}
\end{figure}

The LPBF process repeats the heating and cooling cycles to melt and solidify the laminated powder material.
Furthermore, because this material layer thickness is tens of micrometers, a part-scale analysis is computationally expensive.
To address this, flash heating has been developed to simulate temperature transition by introducing meta-layers scaled up from the actual material layer and applying to each meta-layer a volume heat flux equivalent to laser irradiation \cite{zaeh2010investigations,papadakis2014computational,prabhakar2015computational,li2017efficient,li2018scalable,yang2018prediction,zhang2019resolution,bayat2020part}.
The meta-layer is usually 0.5-1.0 mm.
In other words, scaling up to more than 10 times the actual layer improves computational efficiency with small changes in the computational output of thermal distortion and residual stress \cite{zaeh2010investigations,zhang2019resolution,bayat2020part}.
Based on the above method, we consider a build chamber $\mathit{\Om}$ comprising the part $\mathit{\Om_{c}}$, the support structure $\mathit{\Om_{s}}$, powder $\mathit{\Om_{p}}$, and a build plate $\mathit{\Om_{b}}$, which represents the completed build state as shown in Fig.\ref{fig:buildchamber}.
Here, to represent the intermediate state of the building process, the build chamber is divided into $m$ layers with a fixed thickness in the building direction, as shown in Fig. \ref{fig:analysisdomain}.
The build chamber $\mathit{\Om}$ is defined by each domain $\mathit{\Om}_{i}$ for $1\leq i\leq m$ as follows: 
\begin{equation}
\mathit{\Om} = \mathit{\Om_{1}} \cup{\ldots} \cup{\mathit{\Om_{i}}} \cup{\ldots} \cup{\mathit{\Om_{m}}}.
\label{eq:lpbf1}
\end{equation}
Furthermore, we introduce three subdomains: the active domain $\mathit{\Omega_{A}}$, inactive domain $\mathit{\Omega_{I}}$, and laser irradiation domain $\mathit{\Omega_{L}}$.
The subdomain region depends on the domain number $i$, and each subdomain is defined as: 
 \begin{align}
&\mathit{\Om_{A}} = \mathit{\Om_{1}} \cup{\ldots} \cup{\mathit{\Om_{i}}}, \label{eq:lpbf2}\\
&\mathit{\Om_{I}} = \mathit{\Om} \setminus \mathit{\Om_{A}}, \label{eq:lpbf3}\\
&\mathit{\Om_{L}} = \mathit{\Om_{i}} \setminus{\mathit{\Om_{p}}}. \label{eq:lpbf4}
\end{align}
The subdomain is used to activate layers sequentially from the bottom and apply the volume heat flux to the activated layer, that is, the laser irradiation domain $\mathit{\Omega_{L}}$.
This allowed us to simulate the laser irradiation of each layer during the building process.

Because we focused on the heat flow of the parts after laser irradiation, the analysis model was simplified based on the following assumptions:
First, the part $\mathit{\Omega_{c}}$ and the support structure $\mathit{\Omega_{s}}$ contained in the active domain $\mathit{\Omega_{A}}$ are filled with temperature-independent isotropic bulk materials.
Second, because the thermal conductivity of the powder is significantly smaller than those of the part and the support structure, the powder region $\mathit{\Omega_{p}}$ is negligible \cite{zaeh2010investigations,keller2014new}.
Furthermore, thermal energy loss owing to radiation and convection has also been neglected \cite{roberts2009three,li2016multiscale,soylemez2019thermo}.
Therefore, heat energy is transferred to the build plate by heat conduction via part $\mathit{\Omega_{c}}$ and the support structure $\mathit{\Omega_{s}}$ within the active domain $\mathit{\Omega_{A}}$.
Third, the phase change from powder to solid and latent heat are ignored.
In other words, the laser irradiation domain $\mathit{\Omega_{L}}$ uses the solid material properties during both heating and cooling processes.
Fourth, the boundary $\mathit{\Gamma}$ representing the build plate, which functions as a heat sink, is fixed at a constant temperature $T_{\text {amb}}$ \cite{peng2016part,peng2018fast,zhou2019topology}.
Based on the above assumptions, the transient heat conduction problem that predicts the temperature field $T_{i}(t,\bm{x}):[0,t_{h}]\times\mathit{\Omega_{A}}\rightarrow\mathbb{R}$ in the heating process is governed by the following equation:
\begin{equation}
	\left\{
	\begin{alignedat}{3}
		\hspace{2mm}&\rho c \frac{\partial T_{i}(t,\bm{x})}{\partial t}-\operatorname{div}(k \nabla T_{i}(t,\bm{x}))=q(\bm{x})&\hspace{5mm}&\text { in }&&\left(0, t_{h}\right) \times \mathit{\Omega_{A}}, \\
		\hspace{2mm}&(k \nabla T_{i}(t,\bm{x})) \cdot n=0&\hspace{5mm}&\text { on }&&\left(0, t_{h}\right) \times \partial \mathit{\Omega_{A}} \setminus \mathit{\Gamma}, \\
		\hspace{2mm}&T_{i}(t,\bm{x})=T_{\text {amb}}&\hspace{5mm}&\text {on }&&\left(0, t_{h}\right) \times \mathit{\Gamma}, \\
		\hspace{2mm}&T_{i}(0,\bm{x})=T_{\text {amb}}&\hspace{5mm}&\text {in }&& \mathit{\Omega_{A}},
	\end{alignedat}
\right.
\label{eq:lpbf5}
\end{equation}
for all indices $i = 1,2,\ldots,m$, where $\bm{x}$ represents a point located in the active domain $\mathit{\Omega_{A}}$, $\rho$ is the density, $c$ is the heat capacity, and $k$ is the thermal conductivity.
The volume heat flux $q(\bm{x})$ of the laser irradiation domain $\mathit{\Omega_{L}}$ is defined as follows:
\begin{equation}
	q (\bm{x} ) = \left\{ \begin{array} { l l } { q } & { \text { for } \bm{x} \in \mathit{\Omega_{L}} },\\
	{ 0 } & { \text { otherwise } }.\end{array} \right.
	\label{eq:lpbf6}
\end{equation}
Next, the transient heat conduction problem that predicts the temperature field $T_{i}(t,\bm{x}):[0,t_{c}]\times\mathit{\Omega_{A}}\rightarrow\mathbb{R}$ in the cooling process is governed by the following equation:
\begin{equation}
\left\{
\begin{alignedat}{3}
\hspace{2mm}&\rho c \frac{\partial T_{i}(t,\bm{x})}{\partial t}-\operatorname{div}(k \nabla T_{i}(t,\bm{x}))=0&\hspace{5mm}&\text { in }&&\left(0,t_{c}\right) \times \mathit{\Omega_{A}}, \\
\hspace{2mm}&(k \nabla T_{i}(t,\bm{x})) \cdot n=0&\hspace{5mm}&\text { on }&&\left(0,t_{c}\right) \times \partial \mathit{\Omega_{A}} \setminus \mathit{\Gamma}, \\
\hspace{2mm}&T_{i}(t,\bm{x})=T_{\text {amb}}&\hspace{5mm}&\text {on }&&\left(0,t_{c}\right) \times \mathit{\Gamma}, \\
\hspace{2mm}&T_{i}(0,\bm{x})=T_{i}(t_{h},\bm{x})&\hspace{5mm}&\text {in }&& \mathit{\Omega_{A}},
\end{alignedat}
\right.
\label{eq:lpbf7}
\end{equation}
for all indices $i = 1,2,\ldots,m$.
The next subsection describes an algorithm for predicting the temperature transition of the LPBF building process using the above governing equations.
\subsection{LPBF building process model}
There are two main strategies for activating the layers that represent the building process: the element birth method and the ersatz material approach.
In this study, we apply the ersatz material approach that represents the inactive state by material properties that are $10^{-3}$ smaller than the part and support structure, and then represents the activated state by replacing the ersatz material with the original material properties.

Our LPBF building process algorithm is as follows:
\begin{description}
	\item[Step1.] Inactivate all layers in the chamber domain $\mathit{\Omega}$ divided into $m$ layers.
	\item[Step2.] The domains are activated in sequence from the bottom layer, and the volume heat flux is applied to the activated layer, that is, the laser irradiation domain $\mathit{\Omega_{L}}$.
	\item[Step3.] The temperature field $T_{i}(t,\bm{x})$, defined in Eq. \ref{eq:lpbf5} and \ref{eq:lpbf7} is solved using FEM.
	\item[Step4.] If all layers are activated, the procedure is terminated; otherwise, it returns to the second step.
\end{description}
\subsection{Numerical scheme for the governing equation}
To solve the governing equation using the FEM, Eq. \ref{eq:lpbf5} discretized in space and time is given as follows:
\begin{equation}
\mathbf{C} \frac{\mathbf{T}_{i}^{j}-\mathbf{T}_{i}^{j-1}}{\Delta t^{j}}+\mathbf{K} \mathbf{T}_{i}^{j}=\mathbf{Q}^{j},
\label{eq:lpbf8}
\end{equation}
where $\Delta t^{j}$ is each time step $(j=1:n)$, and $\mathbf{T}_{i}^{j}$ is the temperature vector.
$\mathbf{C}$, $\mathbf{K}$, and $\mathbf{Q}^{j}$ are the heat capacity, conductivity matrices, and volume heat flux vector, respectively.
These are defined using the shape function $\mathbf{N}$ and B-matrix $\mathbf{B}$ as follows:
\begin{align}
	&\mathbf{C}=\int_{\mathit{\Omega}} \mathbf{N}^{\mathsf{T}} \rho_{e} c \mathbf{N} d \Omega, \\
	&\mathbf{K}=\int_{\mathit{\Omega}} \mathbf{B}^{\mathsf{T}} k_{e} \mathbf{B} d \Omega,\\
	&\mathbf{Q}^{j}=\int_\mathit{{\Omega_{L}}} q^{j} \mathbf{N} d \Omega.
	\label{eq:lpbf9}
\end{align}
$\rho_{e}$ and $ k_{e}$ are the values that depend on the domain to which the discretized element $e$ belongs and is defined as follows:
\begin{align}
\rho_{e}= \left\{ \begin{array} { l l } { \rho } & { \text{for} \; e \in \mathit{\Omega_{A}}, } \\ 
{ 10^{-3}\rho } & { \text{for} \; e \in \mathit{\Omega_{I}} }, \end{array} \right.\label{eq:lpbf10}\\
k_{e}= \left\{ \begin{array} { l l } { k } & { \text{for} \; e \in \mathit{\Omega_{A}}, } \\ 
{ 10^{-3}k } & { \text{for} \; e \in \mathit{\Omega_{I}} }. \end{array} \right.\label{eq:lpbf12}
\end{align}
Furthermore, the cooling processes defined in Eqs. \ref{eq:lpbf7} remove the volume heat flux vector $\mathbf{Q}^{j}$ from the above equation.
\subsection{Numerical example for the analytical model}\label{sec:2.4}
\begin{figure}[htbp]
	\begin{center}
		\includegraphics[width=13.0cm]{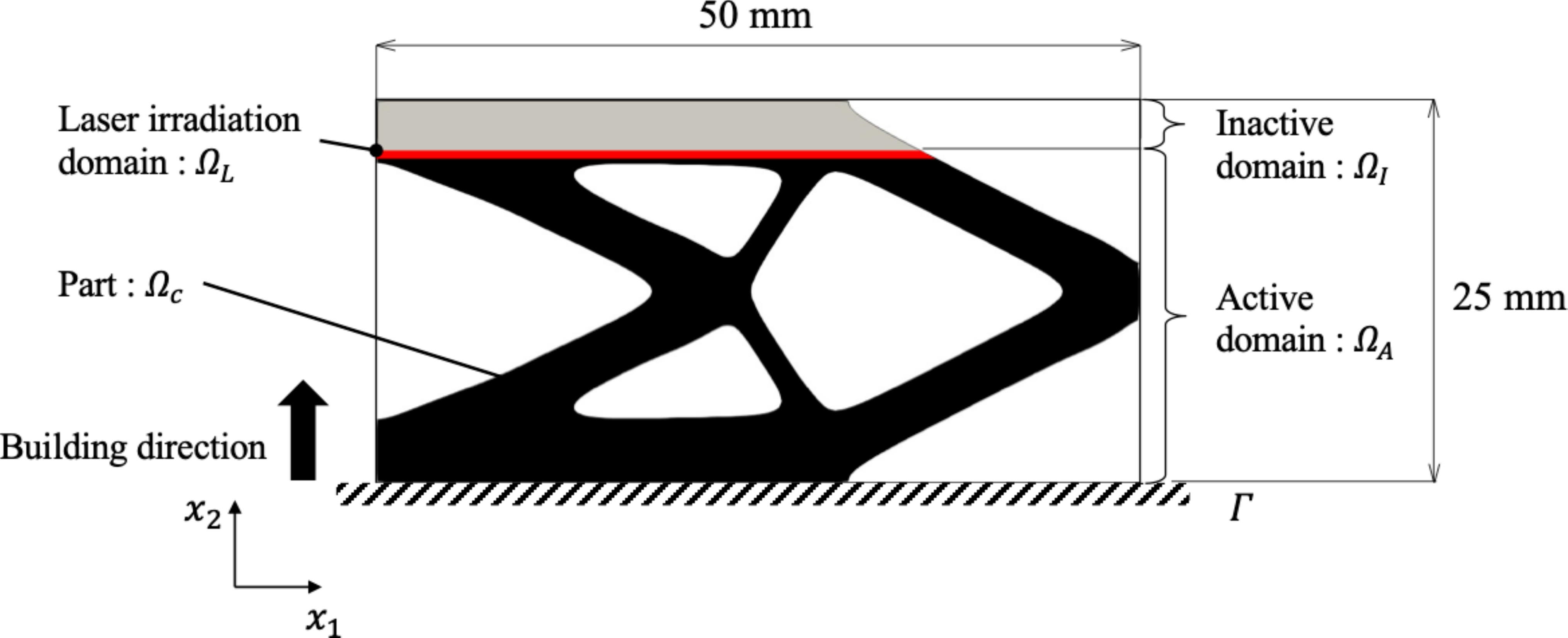}
		\caption{Cantilever model and boundary conditions in the intermediate state of the LPBF building process.}
		\label{fig:numericalex1}
	\end{center}
\end{figure}
This example uses the topology-optimized cantilever model \cite{yamada2010topology}, as shown in Fig. \ref{fig:numericalex1} to evaluate temperature transitions when building an overhanging region.
The material properties and process parameters are listed in Tables \ref{table:param_mat} and \ref{table:param_laser}.
In this paper, we do not focus on the temperature transition in the heating process, so that process is completed in one step.
Then the cooling process begins.
The time step of the cooling process was set to 1 s.
This time step is employed to avoid increasing the computational cost when combined with topology optimization, which requires iterations.
The model is divided in the building direction at 0.5 mm per layer.
This is approximately 10 times the actual material layer.
The effects of layer scaling were investigated by Zhang et al \cite{zhang2019resolution}.
The model is discretized into a mesh of 19,834 second-order triangular elements.
\begin{table}[hbtp]
	\caption{Material properties of AlSi10Mg\cite{alsi10mg}}
	\label{table:param_mat}
	\centering
	\begin{tabular}{ll}
		\hline
		Density $\rho$ (kg/mm$^{3}$)& 2.67$\times10^{-6}$ \\
		Heat capacity $c$ (J/kg K)& 910 \\
		Thermal conductivity $k$ (W/mm K)& 119$\times10^{-3}$ \\
		\hline
	\end{tabular}
\end{table}
\begin{table}[hbtp]
	\caption{Process parameters\cite{li2017efficient}}
	\label{table:param_laser}
	\centering
	\begin{tabular}{ll}
		\hline
		Volume heat flux $q$ (W/mm$^{3}$)& 2$\times10^{4}$ \\
		Heating time per layer $t_{h}$ (s)& 0.5$\times10^{-3}$ \\
		Cooling time per layer $t_{c}$ (s)& 10 \\
		Build plate temperature $T_{amb}$($^\circ$C)& 20\\
		\hline
	\end{tabular}
\end{table}
\begin{figure}[htbp]
	\begin{center}
		\includegraphics[width=13.0cm]{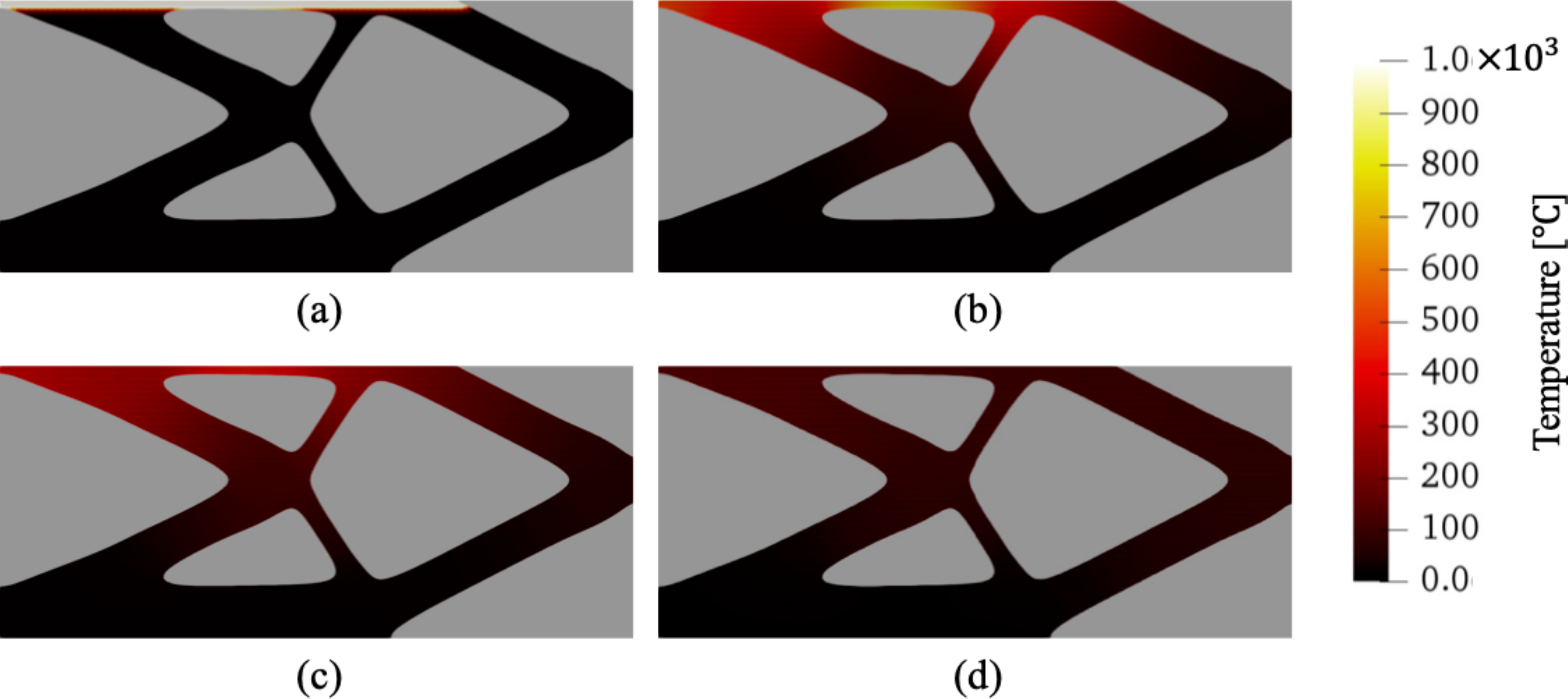}
		\caption{Temperature transition of LPBF process: (a) end of the heating process $t_{h} =$0.5$\times10^{-3}$; (b, c) intermediate time step of the cooling process $t_{c}=$1 s and 2 s; (d) end of the cooling process $t_{c} =$10 s.}
		\label{fig:numericalresult1}
	\end{center}
\end{figure}
\begin{figure}[htbp]
	\begin{center}
		\centering
		\subfigure[]{\includegraphics[width=10.0cm]{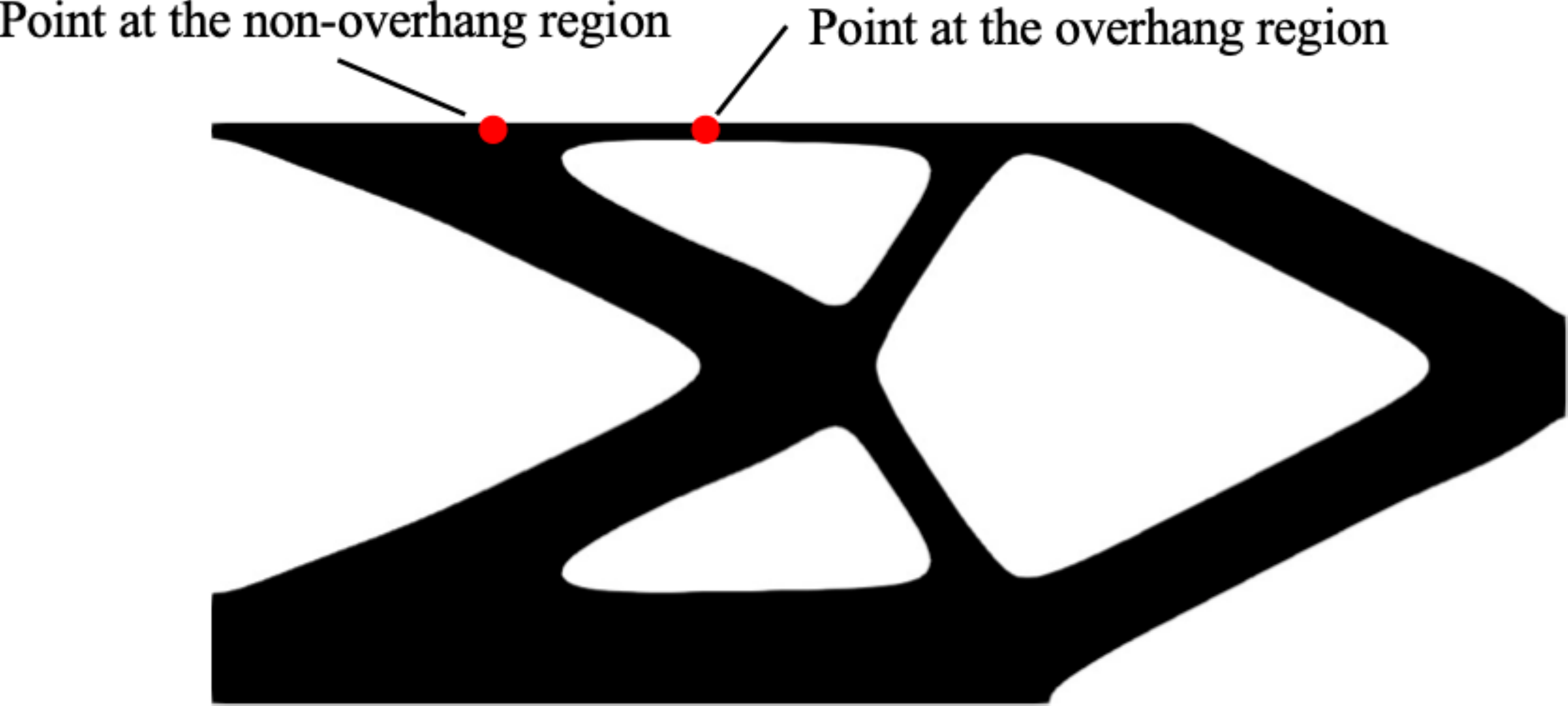}}
		\subfigure[]{\includegraphics[width=10.0cm]{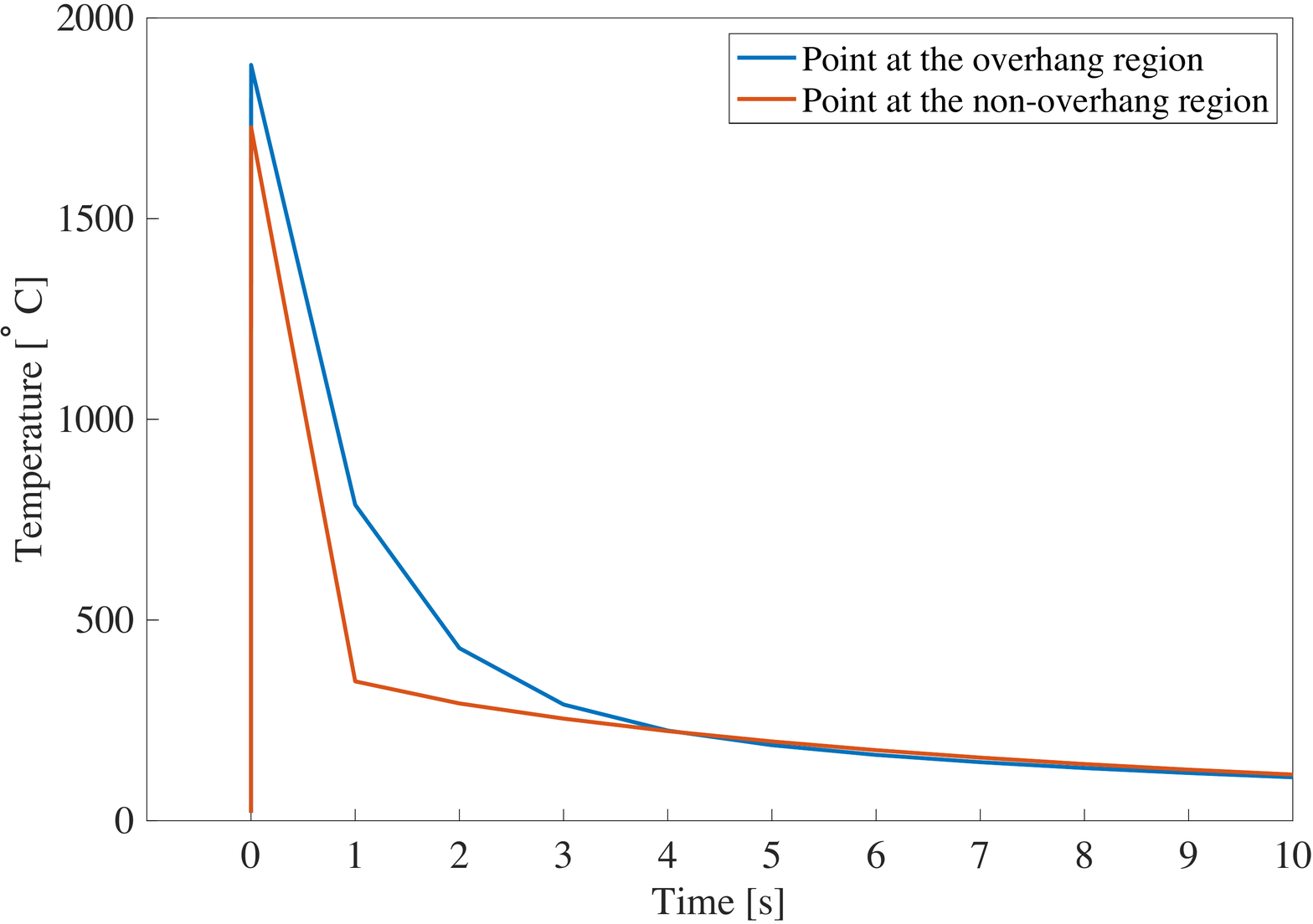}}
		\caption{Comparison of temperature histories in the overhang region and the non-overhang region: (a) Point locations for plotting temperature history (b)Temperature history at each point.}
		\label{fig:temperaturehistory}
	\end{center}
\end{figure}

\begin{figure}[htbp]
	\begin{center}
		\includegraphics[width=9.0cm]{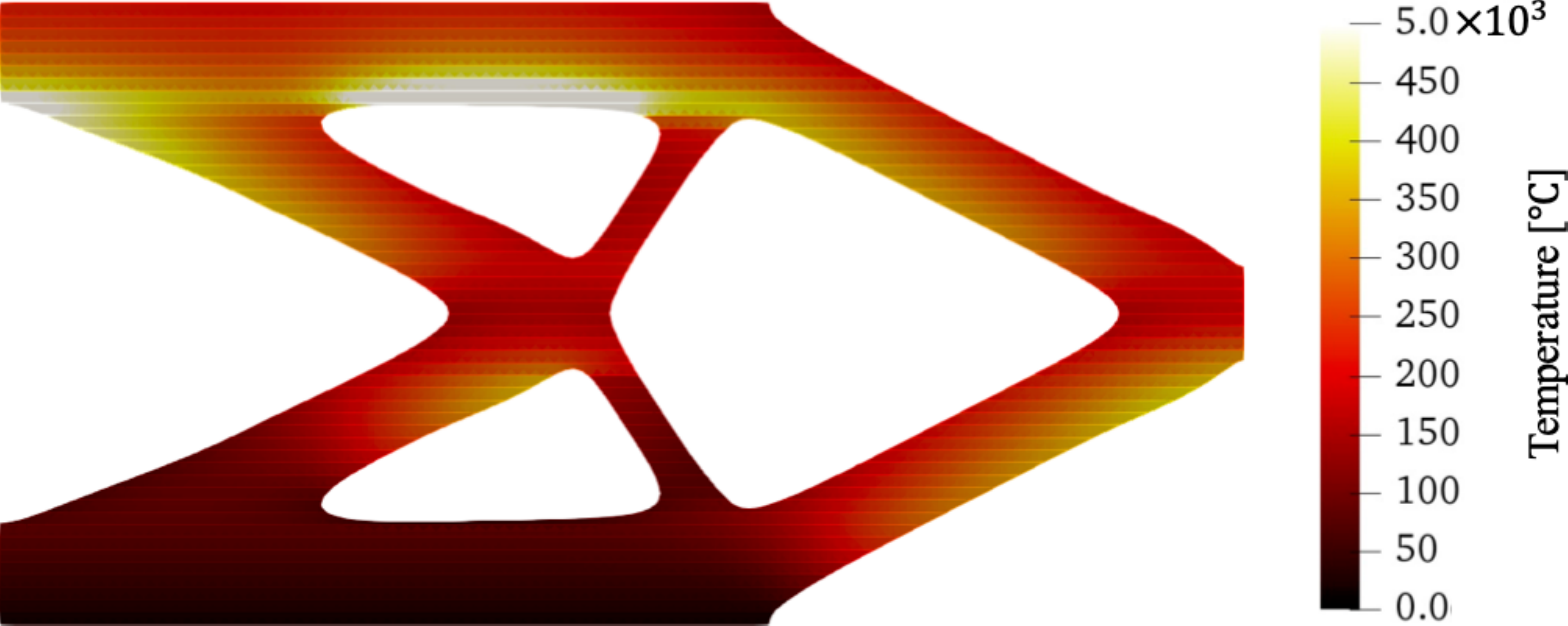}
		\caption{Temperature field of each laser irradiation domain $\mathit{\Omega_{L}}$ in the cooling process $ t_{c}=$1 s.}
		\label{fig:numericalresult2}
	\end{center}
\end{figure}

Figure \ref{fig:numericalresult1} shows the temperature fields at four different time steps in the active domain $\mathit{\Omega_{A}}$.
During the heating process, only the added layer is heated, and almost no heat flows to the lower layer.
In the cooling process, heat energy flows to the lower layer, but if there is a region that blocks the heat flow, such as an overhang, it results in a non-uniform temperature distribution.
The temperature histories of the overhang and non-overhang regions are compared in Fig. \ref{fig:temperaturehistory}.
This result shows that a poor heat dissipation leads to a non-uniform temperature distribution in the cooling process.
Furthermore, after $ t_{c}=$4 s of the cooling process, the cooling rate suddenly decreases and the temperature distribution becomes uniform. 
Therefore, it is necessary to dissipate heat so that the temperature distribution in the added layer becomes uniform within $ t_{c}=$3 s of the cooling process.
Next, Figure \ref{fig:numericalresult2} shows the result of summing the temperature fields of each laser irradiation domain $\mathit{\Omega_{L}}$ in the cooling process $ t_{c}=$1 s. 
This result also shows that the overhang region has the most non-uniform temperature distribution.
Therefore, the support structure must be added at the appropriate location in the void region (the powder $\mathit{\Omega_{p}}$), excluding the parts $\mathit{\Omega_{c}}$ to improve heat dissipation at each layer.
In the following sections, we formulate the support structure optimization problem that maximizes the heat dissipation.
\section{Topology optimization for heat dissipation in LPBF}\label{sec:3}
\subsection{Level-set-based topology optimization}
Topology optimization introduces a fixed design domain and represents the optimal material distribution within that domain by a characteristic function.
In this study, we used a level-set-based topology optimization method \cite{yamada2010topology} to optimize the material distribution in the build chamber $\mathit{\Omega}$, which corresponds to a fixed design domain.
This method uses a level-set function with a value between -1 and 1 to represent the material domain corresponding to the part $\mathit{\Omega_{c}}$ and the support structure $\mathit{\Omega_{s}}$, the void domain corresponding to the powder $\mathit{\Omega_{p}}$, and their boundaries as follows:
\begin{equation}
\left\{ \begin{array} { l l } { 0 < \phi ( \bm{x} ) \leq 1 } & { \text { for } \bm{x} \in  (\mathit{\Omega_{c}} \cup\mathit{\Omega_{s}}) \setminus \partial (\mathit{\Omega_{c}} \cup\mathit{\Omega_{s}}) }, \\
{ \phi ( \bm{x} ) = 0 } & { \text { for } \bm{x} \in \partial (\mathit{\Omega_{c}} \cup\mathit{\Omega_{s}}) }, \\
{ - 1 \leq \phi ( \bm{x} ) < 0 } & { \text { for } \bm{x} \in \mathit{\Omega} \setminus (\mathit{\Omega_{c}} \cup\mathit{\Omega_{s}}) }. \end{array} \right.
\label{eq:lsf1}
\end{equation}
The characteristic function $\chi(\phi)$ is defined using the level-set function as follows:
\begin{equation}
\chi(\phi)= \left\{ \begin{array} { l l } { 1 } & { \text { for } \phi( \bm{x} ) \geq 0 }, \\ { 0 } & { \text { for } \phi( \bm{x} ) < 0 }. \end{array} \right.
\label{eq:lsf2}
\end{equation}
The optimization problem for the optimal material distribution in the build chamber that minimizes the objective function $F$ is formulated as follows:
\begin{equation}
\underset{\phi}{\text{inf}}\;\;\;F(\chi(\phi)).
\label{eq:lsf3}\\
\end{equation}
We replace this optimization problem with the time evolution problem of the level-set function, which introduces fictitious time $s$.
\begin{equation}
\frac { \partial \phi(s) } { \partial s } = - DF',
\label{eq:lsf7}
\end{equation}
where $D$ is a positive parameter, and $F'$ is the derivative of the objective function $F$.
To regularize the above equation, the following Laplacian term is introduced:
\begin{equation}
\frac { \partial \phi(s) } { \partial s } = - D(F'-\tau\nabla^{2}\phi),
\label{eq:lsf8}
\end{equation}
where $\tau>0$ is a parameter that controls the strength of regularization, and increasing $\tau$ leads to a smoother distribution of the level-set functions.
By adjusting $\tau$, regularization was achieved without interfering with the minimization of the objective function.
Furthermore, Eq. \ref{eq:lsf8} discretized in space and time is given as follows:
\begin{equation}
\mathbf{M} \frac{\mathbf{\Phi}(s+\Delta s)-\mathbf{\Phi}(s)}{\Delta s}+\mathbf{Y}\mathbf{\Phi}(s)=\mathbf{D},
\label{eq:lsf9}
\end{equation}
where $\mathbf{\Phi}$ is the vector of the level-set function, and $\mathbf{M}$, $\mathbf{Y}$, and $\mathbf{D}$ are described as follows:
\begin{align}
&\mathbf{M}=\int_{\mathit{\Omega}} \mathbf{N}^{\mathsf{T}} \mathbf{N} d \Omega, \\
&\mathbf{Y}=\int_{\mathit{\Omega}} \mathbf{B}^{\mathsf{T}} D \tau \mathbf{B} d \Omega,\\
&\mathbf{D}=\int_\mathit{{\Omega}} DF' \mathbf{N} d \Omega.
\end{align}
\subsection{Formulation of optimization problem}
In this study, the objective function $F$ minimizes the squared error between the temperature of the part $\mathit{\Omega_{c}}$ contained in each laser irradiation domain $\mathit{\Omega_{L}}$ and the build plate temperature $\mathbf{T}_{amb}$,　which corresponds to maximizing heat dissipation,　which corresponds to maximizing heat dissipation during the cooling process.
\begin{equation}
	F=\sum_{i=1}^{m}\sum_{j=1}^{n}\left(\mathbf{T}_{i}^{j} - \mathbf{T}_{amb}\right)^{2}\Delta t^{j}\hspace{5mm}\text{for} \; e \in \mathit{\Omega_{c}}\cap\mathit{\Omega_{L}},
	\label{eq:op1}
\end{equation}
Thus, the optimization problem for determining the optimal configuration of the support structure $\mathit{\Omega_{s}}$ to maximize the heat dissipation of part $\mathit{\Omega_{c}}$ under volume constraints can be formulated as follows:
\begin{equation}
\begin{split}
\inf_{\mathbf{\Phi}}\hspace{17mm} &F,&\\
\text{subject to}: \hspace{2mm}&G=\int_{\mathit{\Omega}\setminus\mathit{\Omega_{c}}}\chi \hspace{1mm}d\Omega-V_\text{max}\leq 0,&\\
&\mathbf{R}_{i}^{j}=\mathbf{C} \frac{\mathbf{T}_{i}^{j}-\mathbf{T}_{i}^{j-1}}{\Delta t^{j}}+\mathbf{K} \mathbf{T}_{i}^{j}=\bm{0},&\\
\label{eq:op2}
\end{split}
\end{equation}
for all indices $j = 1,2,\ldots,n$, and $i = 1,2,\ldots,m$. Here, $G$ represents the volume constraint, and $V_{max}$ is the upper limit of the material volume in the chamber, excluding the part $\mathit{\Omega}\setminus\mathit{\Omega_{c}}$.
$\mathbf{R}_{i}^{j}$ represents the governing equation in the cooling process.
\subsection{Sensitivity analysis}\label{sec:3.3}
The sensitivity of the objective function $F$ in Eq. \ref{eq:op1} is derived using the adjoint variable method.
First, by introducing an adjoint variable $\bm{\lambda}_{i}^{j}$ into the governing equation, the extended objective function $\tilde F$ can be written as
 \begin{equation}
	\tilde F=F+\sum_{i=1}^{m}\sum_{j=1}^{n} \bm{\lambda}_{i}^{j\mathsf{T}}\mathbf{R}_{i}^{j}.
	\label{eq:op3}
\end{equation}
Next, the above extended objective function is differentiated by the design variable $\mathbf{\Phi}$.
\begin{equation}
\begin{aligned}
\frac{\partial \tilde{F}}{\partial \mathbf{\Phi}}=& \frac{\partial F}{\partial \mathbf{\Phi}}
+\sum_{i=1}^{m}\sum_{j=1}^{n} \frac{\partial F}{\partial \mathbf{T}_{i}^{j}} \frac{\partial \mathbf{T}_{i}^{j}}{\partial \mathbf{\Phi}} \\
&+\sum_{i=1}^{m}\sum_{j=1}^{n} \bm{\lambda}_{i}^{j\mathsf{T}}\left(\frac{\partial \mathbf{R}_{i}^{j}}{\partial \mathbf{\Phi}}
+\frac{\partial \mathbf{R}_{i}^{j}}{\partial \mathbf{T}_{i}^{j}} \frac{\partial \mathbf{T}_{i}^{j}}{\partial \mathbf{\Phi}}
+\frac{\partial \mathbf{R}_{i}^{j}}{\partial \mathbf{T}_{i}^{j-1}} \frac{\partial \mathbf{T}_{i}^{j-1}}{\partial \mathbf{\Phi}}\right).
	\label{eq:op4}
\end{aligned}
\end{equation}
The above equation can be rearranged as follows:
\begin{equation}
\begin{aligned}
\frac{\partial \tilde{F}}{\partial \mathbf{\Phi}}=& \frac{\partial F}{\partial \mathbf{\Phi}}
+\sum_{i=1}^{m}\sum_{j=1}^{n}\bm{\lambda}_{i}^{j\mathsf{T}}\frac{\partial \mathbf{R}_{i}^{j}}{\partial \mathbf{\Phi}}
+\sum_{i=1}^{m}\left(\frac{\partial F}{\partial \mathbf{T}_{i}^{n}}
+\bm{\lambda}_{i}^{n\mathsf{T}}\frac{\partial \mathbf{R}_{i}^{n}}{\partial \mathbf{T}_{i}^{n}} \right)
\frac{\partial \mathbf{T}_{i}^{n}}{\partial \mathbf{\Phi}} \\
&+\sum_{i=1}^{m}\sum_{j=1}^{n-1}\left(\frac{\partial F}{\partial \mathbf{T}_{i}^{j}}
+\bm{\lambda}_{i}^{j\mathsf{T}}\frac{\partial \mathbf{R}_{i}^{j}}{\partial \mathbf{T}_{i}^{j}}
+\bm{\lambda}_{i}^{j+1\mathsf{T}}\frac{\partial \mathbf{R}_{i}^{j+1}}{\partial \mathbf{T}_{i}^{j}}\right)
\frac{\partial \mathbf{T}_{i}^{j}}{\partial \mathbf{\Phi}}.
	\label{eq:op5} 
\end{aligned}
\end{equation}
From the above, the adjoint equation in the time step $n$ is described as
\begin{equation}
\frac{\partial F}{\partial \mathbf{T}_{i}^{n}}
+\bm{\lambda}_{i}^{n\mathsf{T}}\frac{\partial \mathbf{R}_{i}^{n}}{\partial \mathbf{T}_{i}^{n}}
=\bm{0}.
	\label{eq:op6}
\end{equation}
The adjoint variable $\bm{\lambda}_{i}^{n}$ can be obtained from the transient heat equation in Eq. \ref{eq:lpbf7} at the time step $n$.
Furthermore, the adjoint equation of the time step $1 \leq j\leq n-1$ is described as follows:
\begin{equation}
\frac{\partial F}{\partial \mathbf{T}_{i}^{j}}
+\bm{\lambda}_{i}^{j\mathsf{T}}\frac{\partial \mathbf{R}_{i}^{j}}{\partial \mathbf{T}_{i}^{j}}
+\bm{\lambda}_{i}^{j+1\mathsf{T}}\frac{\partial \mathbf{R}_{i}^{j+1}}{\partial \mathbf{T}_{i}^{j}}
=\bm{0},
	\label{eq:op7}
\end{equation}
where $\frac{\partial F}{\partial \mathbf{T}_{i}^{j}}$, $\frac{\partial \mathbf{R}_{i}^{j}}{\partial \mathbf{T}_{i}^{j}}$ and $\frac{\partial \mathbf{R}_{i}^{j+1}}{\partial \mathbf{T}_{i}^{j}}$ are calculated as follows:
\begin{align}
&\frac{\partial F}{\partial \mathbf{T}_{i}^{j}}=2\left(\mathbf{T}_{i}^{j} - \mathbf{T}_{amb}\right)\Delta t^{j}\\
&\frac{\partial \mathbf{R}_{i}^{j}}{\partial \mathbf{T}_{i}^{j}}=\frac{1}{\Delta t^{j}}\mathbf{C} +\mathbf{K},\\
&\frac{\partial \mathbf{R}_{i}^{j+1}}{\partial \mathbf{T}_{i}^{j}}=-\frac{1}{\Delta t^{j}}\mathbf{C}.
\end{align}
The adjoint variable of time step $1 \leq j\leq n-1$ can be obtained by solving the above equation in the backward direction with $\bm{\lambda}_{i}^{n}$ as the initial condition.
By substituting the adjoint variables obtained from Eqs. \ref{eq:op6}  and \ref{eq:op7}, the sensitivity of the objective function can be described as follows:
\begin{equation}
\frac{\partial \tilde{F}}{\partial \mathbf{\Phi}}= \frac{\partial F}{\partial \mathbf{\Phi}}
+\sum_{i=1}^{m}\sum_{j=1}^{n}\bm{\lambda}_{i}^{j\mathsf{T}}\frac{\partial \mathbf{R}_{i}^{j}}{\partial \mathbf{\Phi}},
\label{eq:op8} 
\end{equation}
where $\frac{\partial \mathbf{R}_{i}^{j}}{\partial \mathbf{\Phi}}$ is calculated as:
\begin{equation}
\begin{aligned}
\frac{\partial \mathbf{R}_{i}^{j}}{\partial \mathbf{\Phi}}=&
\left(\frac{1}{\Delta t^{j}} \frac{\partial \mathbf{C}}{\partial \mathbf{\Phi}}+ \frac{\partial \mathbf{K}}{\partial \mathbf{\Phi}}\right) \mathbf{T}_{i}^{j}
-\left(\frac{1}{\Delta t^{j}} \frac{\partial \mathbf{C}}{\partial \mathbf{\Phi}}\right)
\mathbf{T}_{i}^{j-1}.
\end{aligned}
\end{equation}
Furthermore, because the objective function does not depend on the design variable, it becomes:
\begin{equation}
\frac{\partial F}{\partial \mathbf{\Phi}}=\bm{0}.
\end{equation}
Because the volume heat flux is applied to the support structure contained in the laser irradiation domain $\mathit{\Omega_{L}}$, it does not contribute to heat dissipation.
Therefore, the laser irradiation domain $\mathit{\Omega_{L}}$ is not included in the design sensitivity.
\section{Numerical implementation}\label{sec:4}
\subsection{Optimization algorithm}\label{sec:4.1}
The optimization algorithm is as follows.
\begin{description}
	\item[Step1.] The initial value of the level-set function was set.
	\item[Step2.] The temperature field $\mathbf{T}_{i}^{j}$ defined in Eqs. \ref{eq:lpbf5} and \ref{eq:lpbf7} is solved according to the LPBF building process algorithm.
	\item[Step3.] The objective function $F$, defined in Eq. \ref{eq:op1} is evaluated using the solution of the transient heat conduction problem.
	\item[Step4.] If the change ratio of the objective function is less than 0.01$\%$ in 5 consecutive iterations and the volume constraint is satisfied, it is assumed that convergence is established and the optimization procedure is terminated; otherwise, the adjoint variables $\bm{\lambda}_{i}^{j}$ defined in Eqs. \ref{eq:op6} and \ref{eq:op7} are solved using FEM, and the sensitivity of the objective function $F$ is calculated using Eq. \ref{eq:op8}.
	\item[Step5.] The level-set function is updated using Eq. \ref{eq:lsf8} based on sensitivity; then the procedure returns to the second step.
\end{description}
\subsection{Regularization of the boundary between the two domains}
In the FEM analysis, generating a mesh along the boundary between the material and void domains for each optimization iteration increases the computational cost.
In this study, the boundary is expressed by approximating the characteristic function \cite{allaire2004structural} without generating a mesh.
We assume that the void domain has smaller material properties than the material domain, and the material properties at the boundary change smoothly.
Each material property $\rho$ and $k$ in the active domain $\mathit{\Omega_{A}}$ uses the extended material properties expressed by the following equations:
\begin{align}
\tilde{\rho}(\phi;w) = \left\{(1-d)H(\phi;w)+d\right\}\rho,\label{eq:ni1}\\
\tilde{k}(\phi;w) = \left\{(1-d)H(\phi;w)+d\right\}k,\label{eq:ni2}
\end{align}
where $H(\phi;w)$ is defined as:
\begin{equation}
H(\phi;w) := \left\{\begin{array}{ll}
1 & \text { for } \phi>w, \\
\frac{1}{2}+\frac{\phi}{w}\left(\frac{15}{16}-\frac{\phi^{2}}{w^{2}}\left(\frac{5}{8}-\frac{3}{16} \frac{\phi^{2}}{w^{2}}\right)\right) & \text { for }-w \leq \phi \leq w, \\
0 & \text { for } \phi<-w,
\end{array}\right.\label{eq:ni3}
\end{equation}
where $w$ represents the width of the transition and $d$ is the coefficient of the material properties for the void domains.
\section{Numerical examples for the support optimization}\label{sec:5}
This section demonstrates the effectiveness and validity of the proposed optimization method for the support structure to maximize heat dissipation through 2D and 3D numerical examples.
\subsection{Benchmark design examples}\label{sec:5.1}
\begin{figure}[htbp]
	\begin{center}
		\centering
		\subfigure[]{\includegraphics[width=10.0cm]{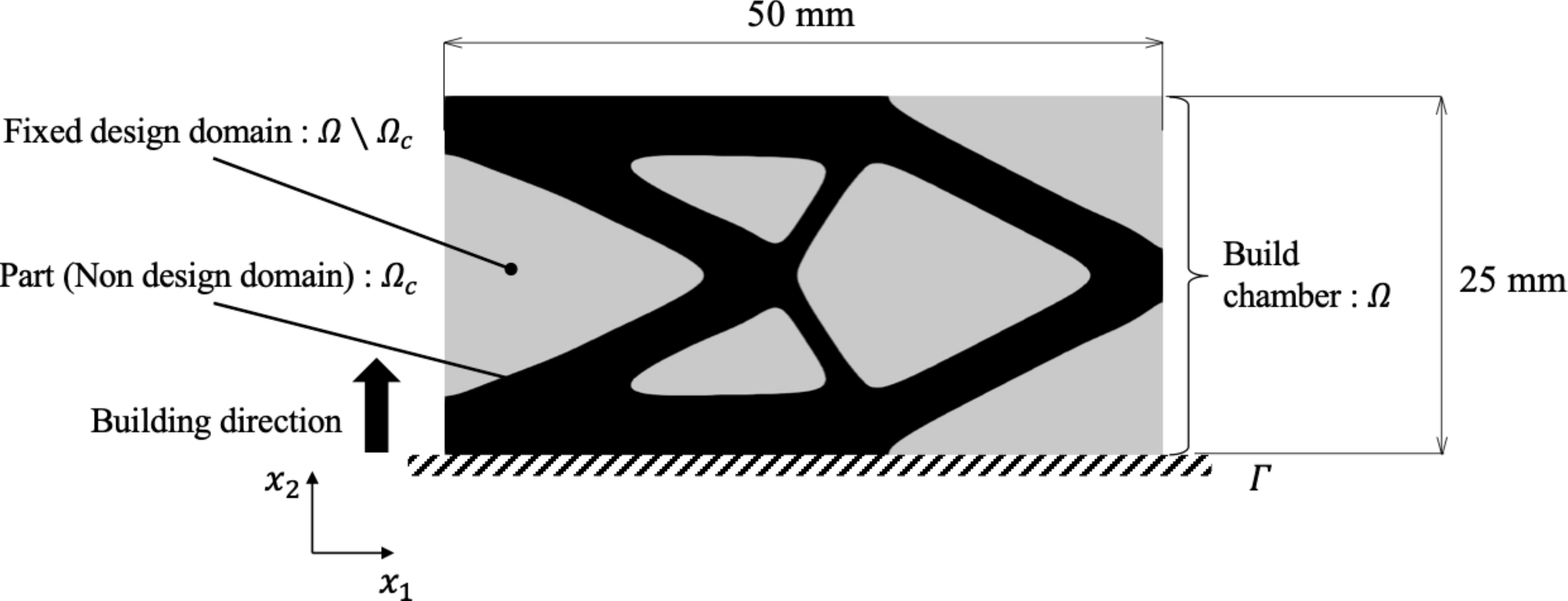}}
		\subfigure[]{\includegraphics[width=13.0cm]{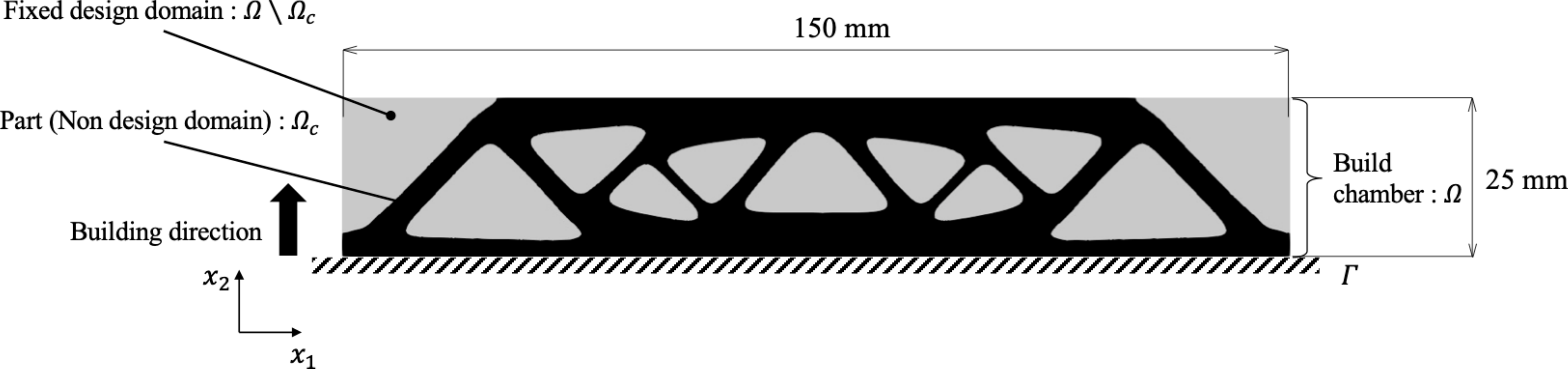}}
		\subfigure[]{\includegraphics[width=13.0cm]{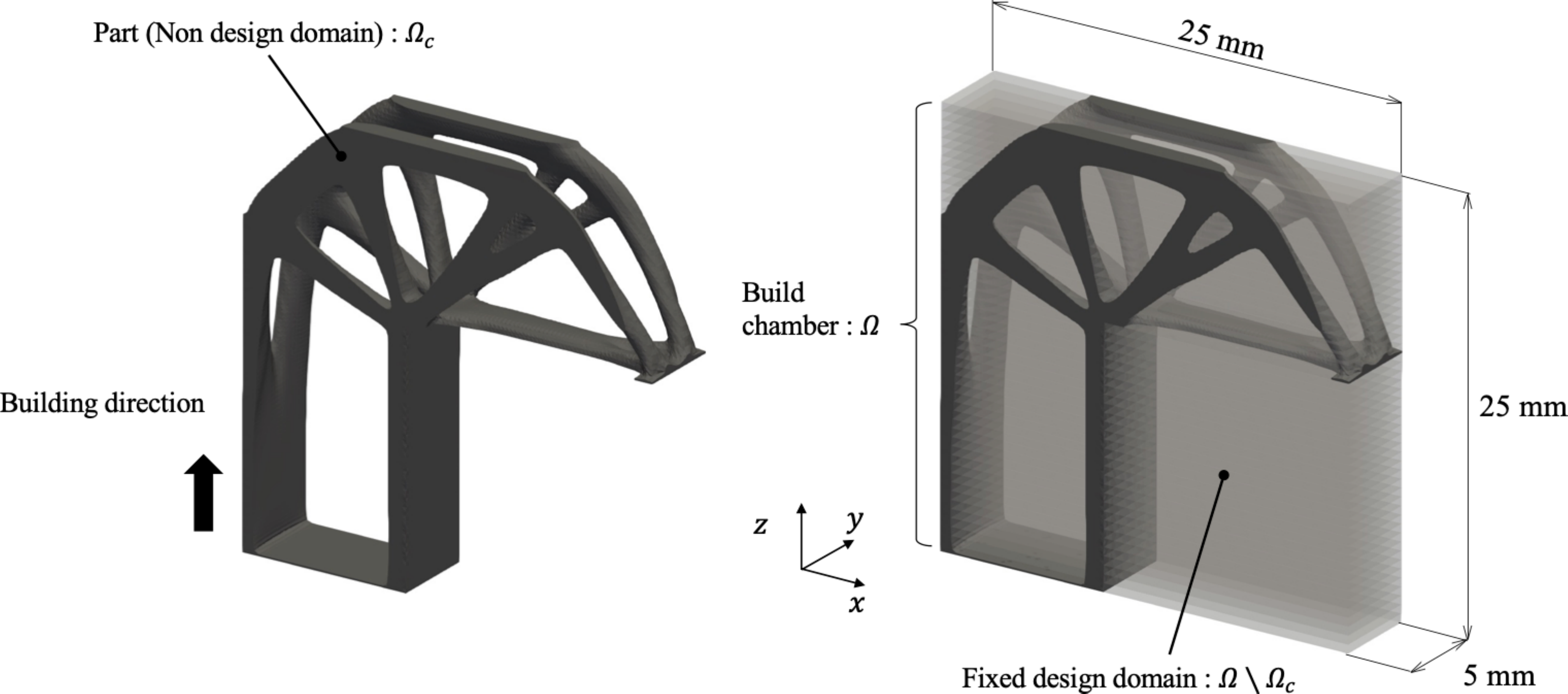}}
		\caption{Problem setting: (a)cantilever model; (b)MBB beam model; (c) L-bracket model.}
		\label{fig:NumericalExample}
	\end{center}
\end{figure}

We consider the optimal support structure using the optimized 2D cantilever, MBB beam, and 3D L-bracket models, as shown in Fig.  \ref{fig:NumericalExample} as the part $\mathit{\Omega{c}}$.
Black and gray represent the non-design and fixed design domains, respectively.
Both build chambers are divided into $m = 50$ layers with a layer thickness of 0.5 mm in the building direction.
The meshes of the cantilever and MBB beam models comprised 43,584 and 64,310 second-order triangular elements, respectively.
The L-bracket model was discretized into a mesh of 1,413,753 second-order tetrahedral elements.
From the result of Subsection \ref{sec:2.4}, the time step $n$ was set to 3 $(j = 1: 3)$ in Eq. \ref{eq:op1}.
The material properties of the part and support structure as well as the boundary conditions are listed in Tables \ref{table:param_mat} and \ref{table:param_laser}.
In order to have the volume the same as that of conventional support structure to compared to later, the upper limit of the material volume was set to 21\% for the cantilever model, 17.4\% for the MBB beam model, and 20\% for the L-bracket model.
The regularization parameter $\tau$ was set to 1 $\times$ $10^{-4}$.
The parameter $D$ in Eq. \ref{eq:lsf8} was set to 0.8, and the parameters $w$ and $d$ in Eq. \ref{eq:ni3} were set to 0.9 and 1 $\times$ $10^{-3}$, respectively.

\begin{figure}[htbp]
	\begin{center}
	\centering
	\subfigure[]{\includegraphics[width=6.0cm]{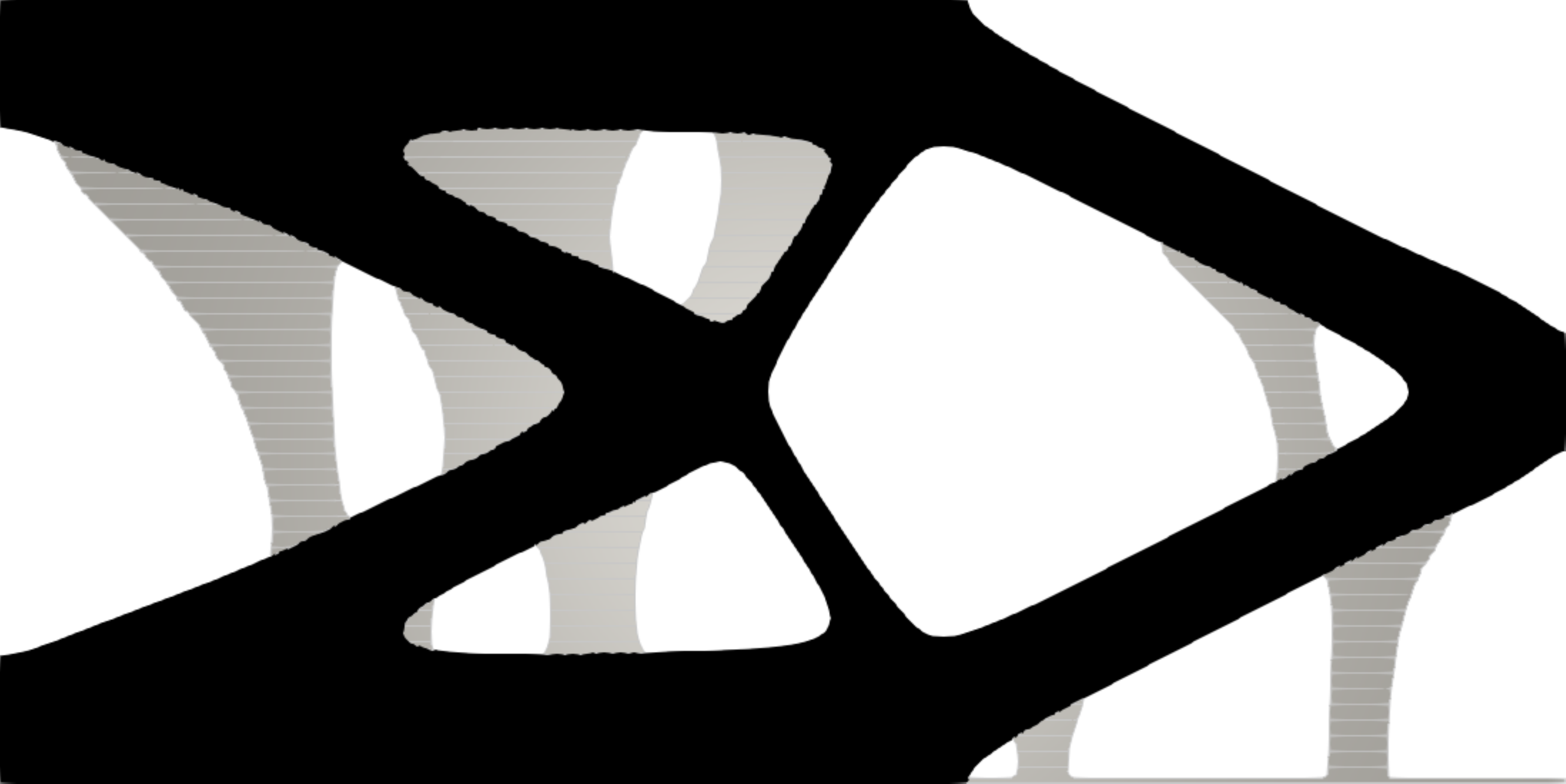}}
	\subfigure[]{\includegraphics[width=9.0cm]{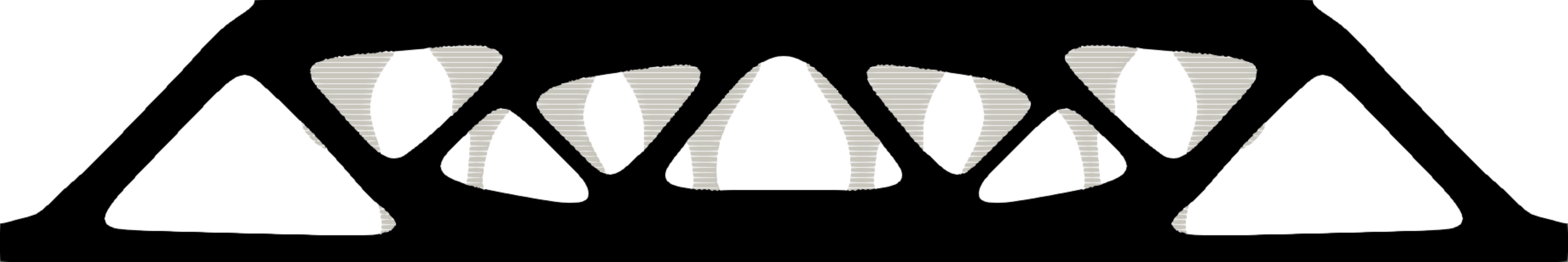}}
	\subfigure[]{\includegraphics[width=13.0cm]{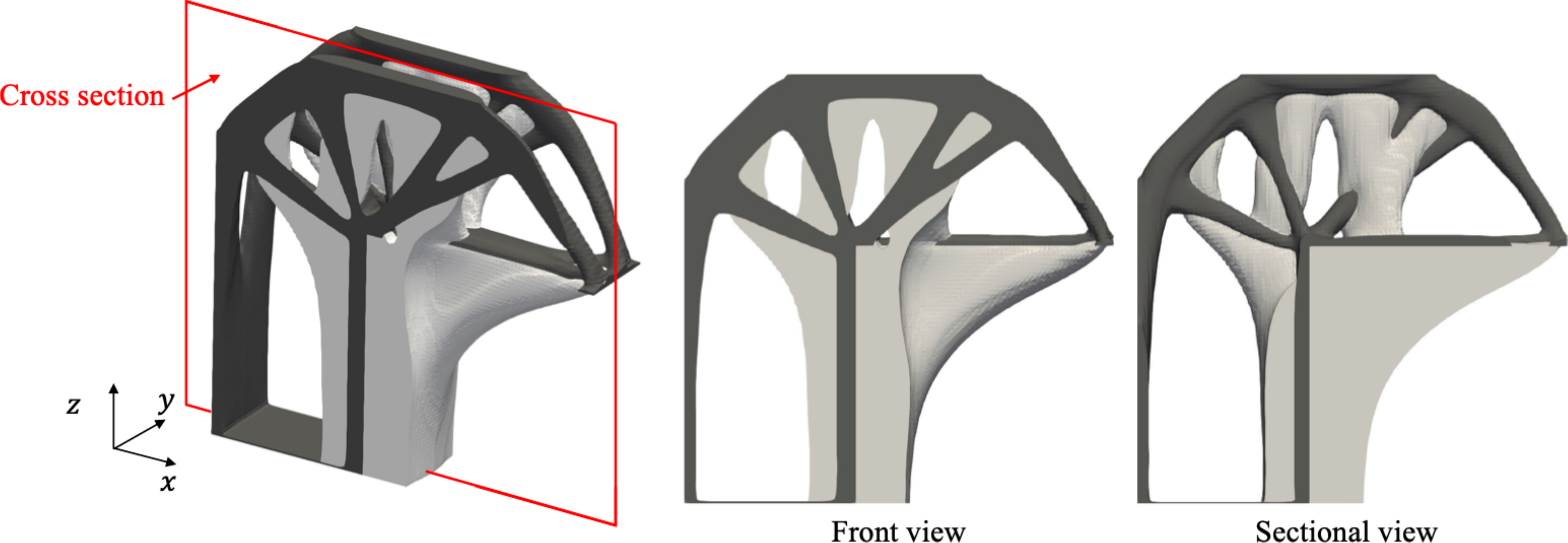}}	
	\caption{Optimal configurations: (a)cantilever model; (b)MBB beam model; (c) L-bracket model. }
	\label{fig:result_unsteady}
\end{center}
\end{figure}
\begin{figure}[htbp]
	\begin{center}
		\subfigure[]{\includegraphics[width=7.0cm]{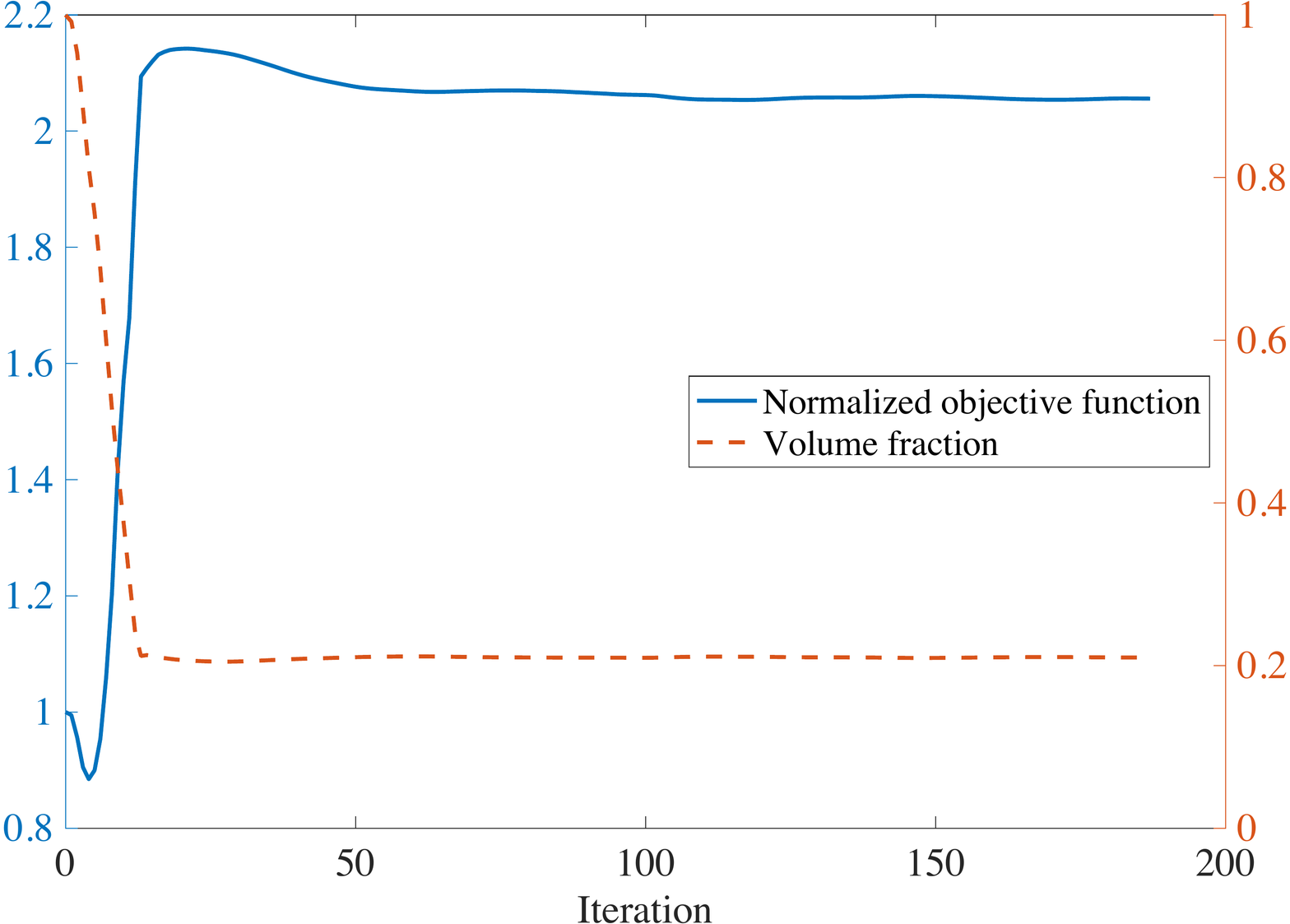}}
		\subfigure[]{\includegraphics[width=7.0cm]{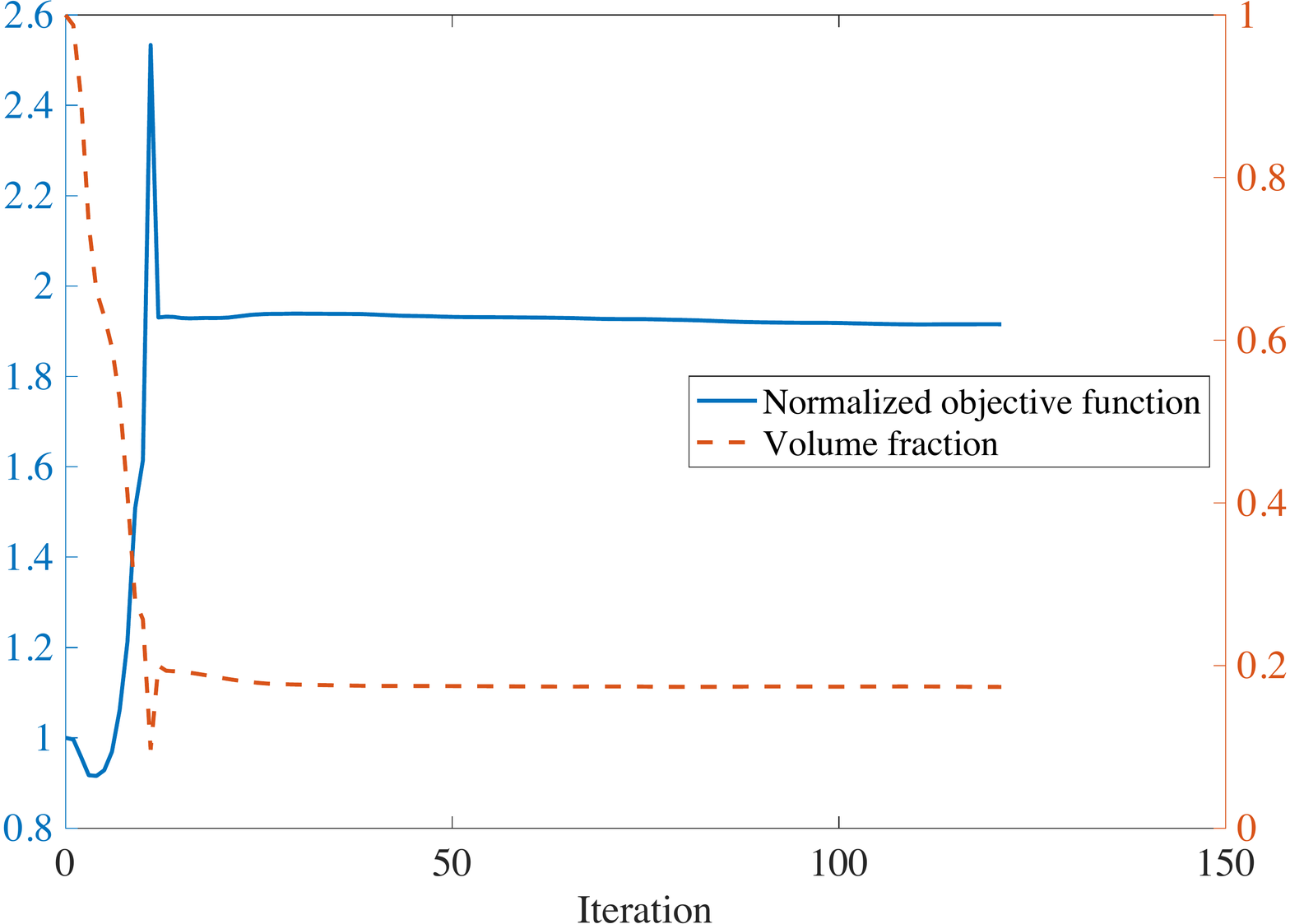}}
		\subfigure[]{\includegraphics[width=7.0cm]{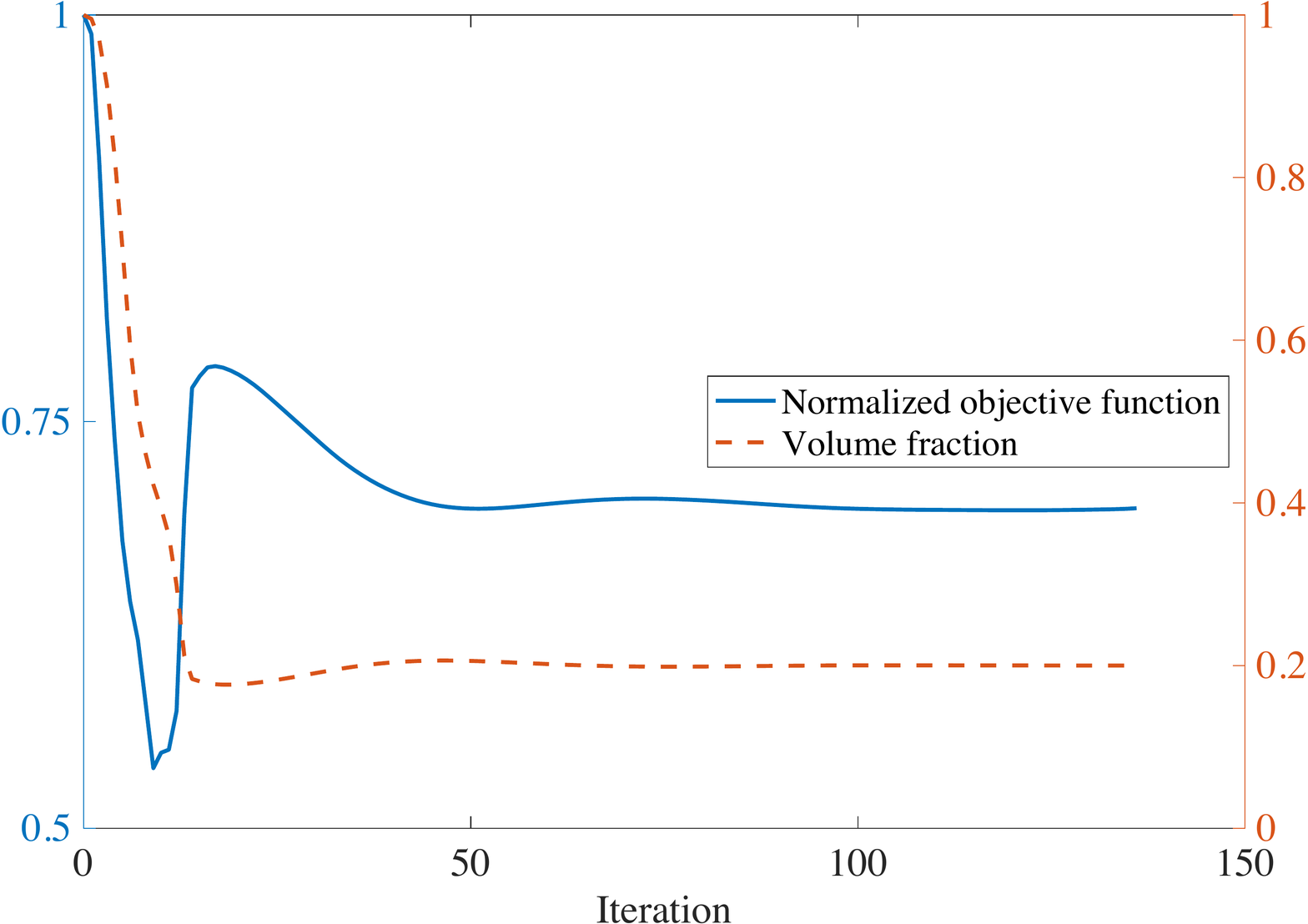}}
		\caption{Convergence history of the objective function and volume constraint: (a)cantilever model; (b)MBB beam model; (c)L-bracket model.}
		\label{fig:obj_vol}
	\end{center}
\end{figure}
\begin{figure}[htbp]
	\begin{center}
		\includegraphics[width=14.0cm]{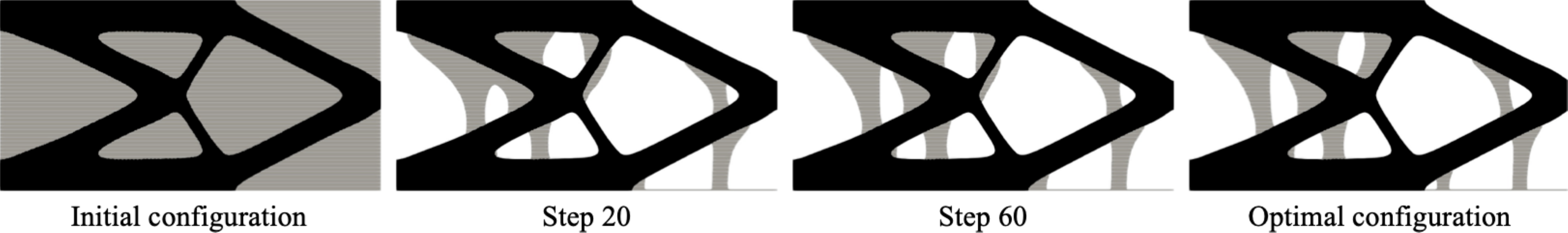}
		\caption{Initial configuration, intermediate results and optimal configuration of cantilever model.}
		\label{fig:Intermidiate}
	\end{center}
\end{figure}
Figure \ref{fig:result_unsteady} shows the optimal configurations for each model.
In all the results, it can be seen that the support structure is added to the overhang region, where the heat dissipation is poor.
	Figure \ref{fig:obj_vol} shows the convergence history of the objective function and volume constraint for each model, Figure \ref{fig:Intermidiate} shows the Initial , intermediate, and optimal configuration of the cantilever model representing each model.
As the number of iterations increases, the objective function decreases while satisfying the volume constraint.
The computational time until convergence was 4 h for the cantilever model, 2.5 h for the MBB beam model, and 34 h for the L-bracket model.
The calculations of the 2D models were run on 14 Intel Xeon E5-2687W cores, and the 3D model was run on 28 Intel Xeon E5-2687W cores.
\subsection{Comparison with conventional support structure}
This subsection examines the effectiveness and validity of the optimized support structure.
Specifically, we compare the optimized support with the traditional support for three items: the temperature field when building the overhang region, the sum of the temperature fields of each laser irradiation domain $\mathit{\Omega_{L}}$, and the objective function.
\begin{figure}[htbp]
	\begin{center}
		\subfigure[]{\includegraphics[width=13.0cm]{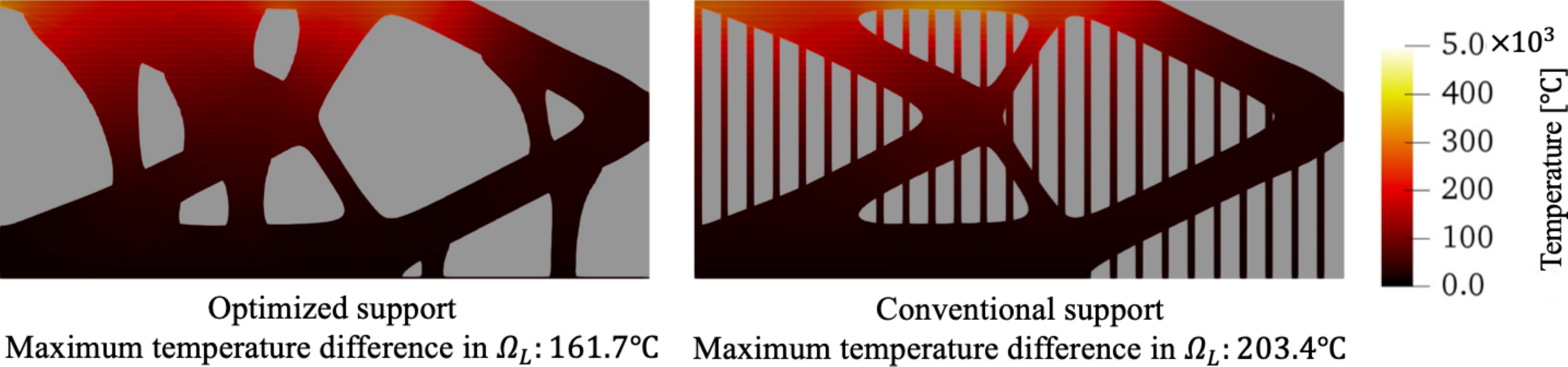}}
		\subfigure[]{\includegraphics[width=12.0cm]{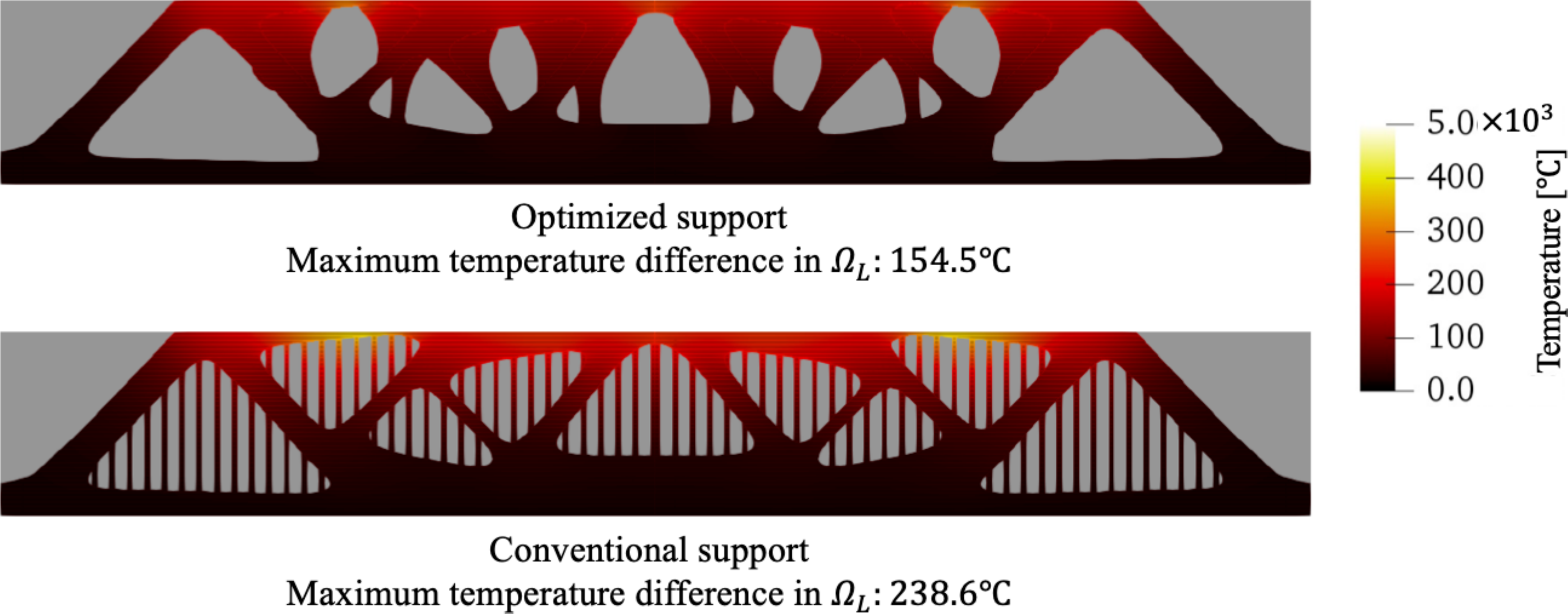}}
		\subfigure[]{\includegraphics[width=12.0cm]{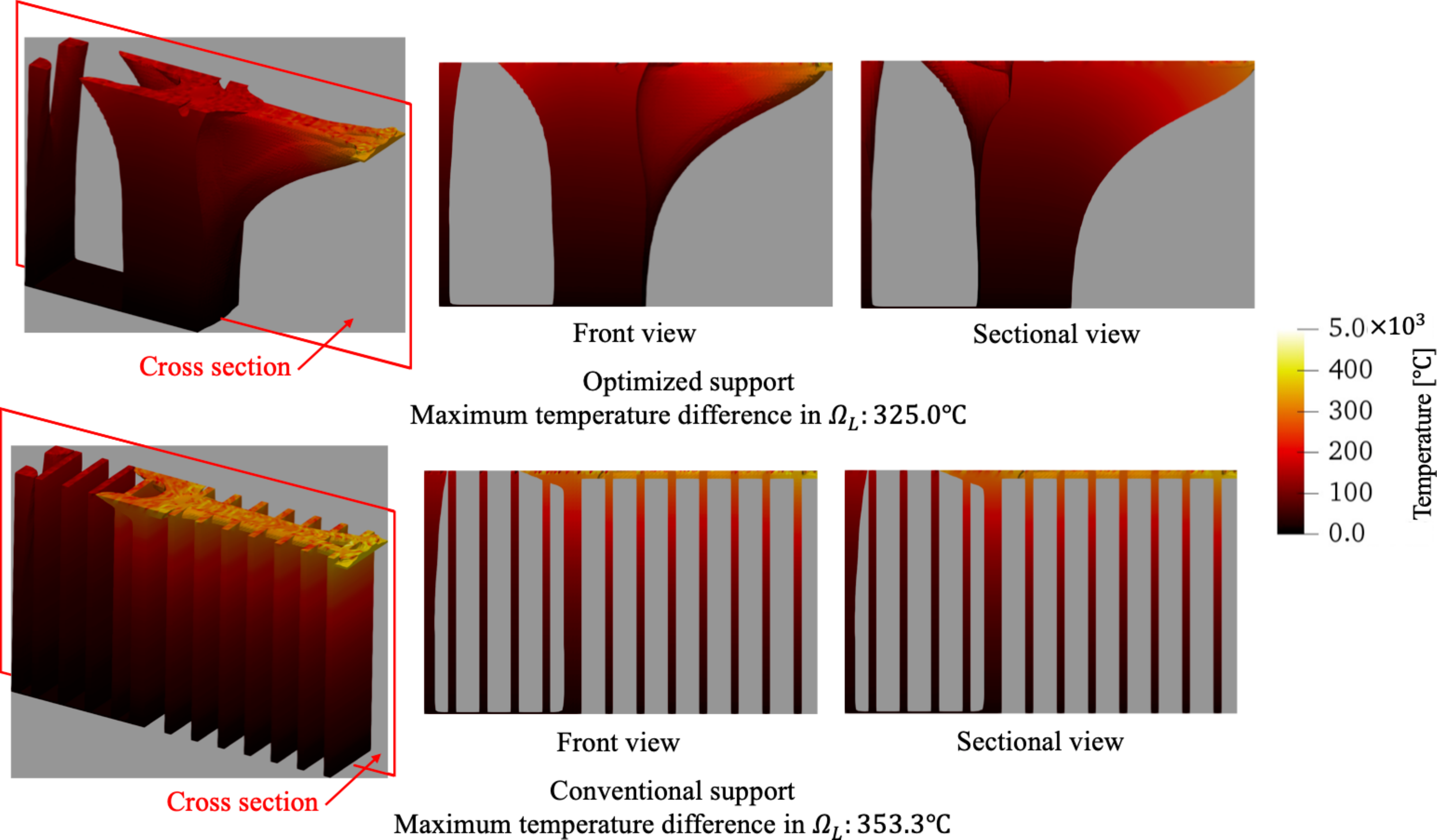}}
		\caption{Temperature field in the cooling process $ t_{c}=$1 s when building the overhang region: (a)cantilever model; (b)MBB beam model; (c)L-bracket model.}
		\label{fig:tc1}
	\end{center}
\end{figure}
\begin{figure}[htbp]
	\begin{center}
		\subfigure[]{\includegraphics[width=13.0cm]{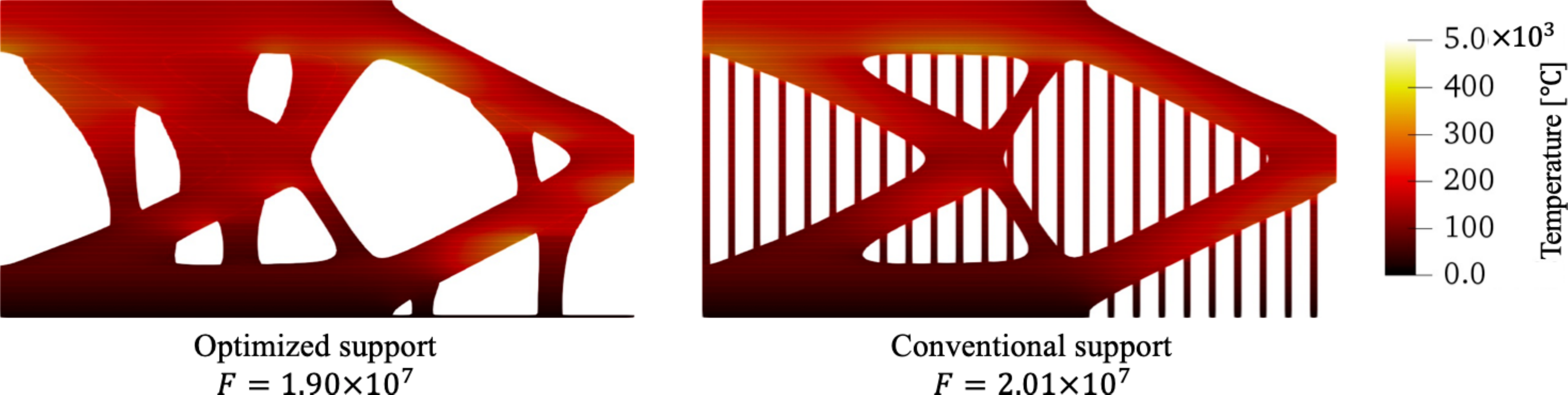}}
		\subfigure[]{\includegraphics[width=12.0cm]{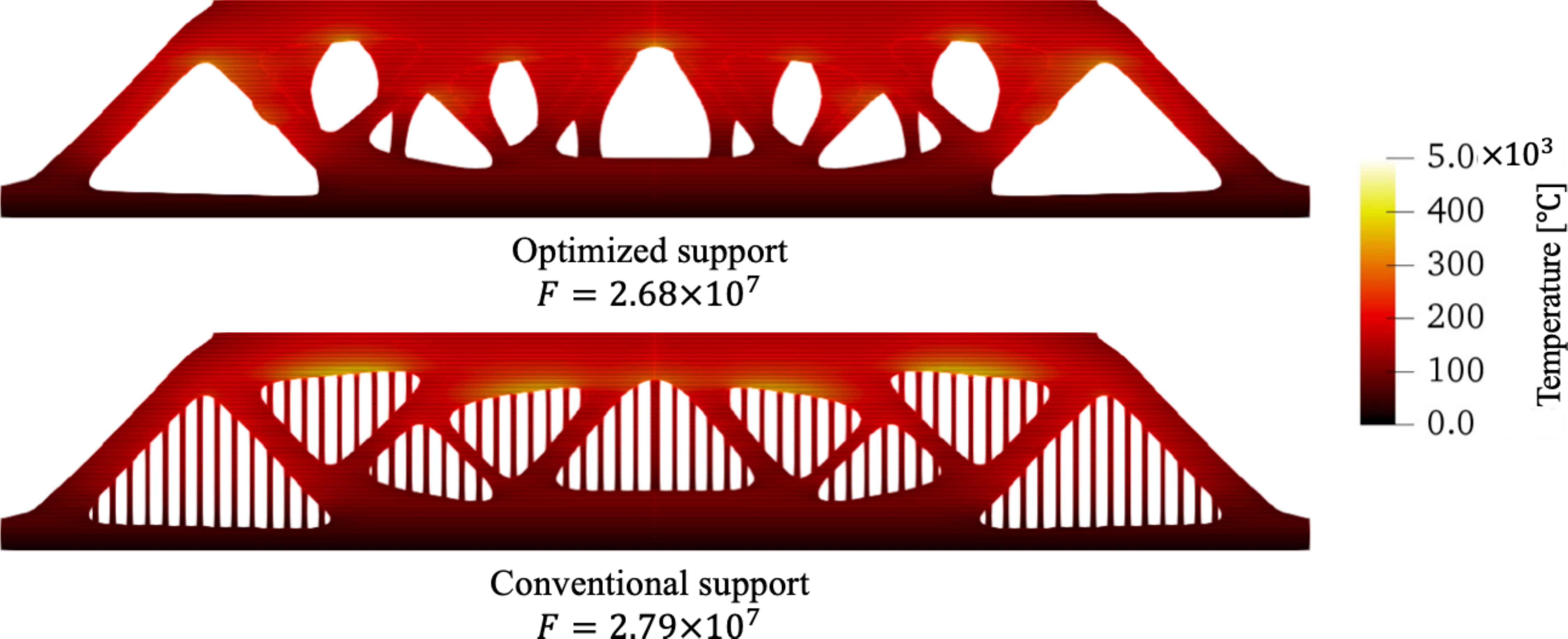}}
		\subfigure[]{\includegraphics[width=12.0cm]{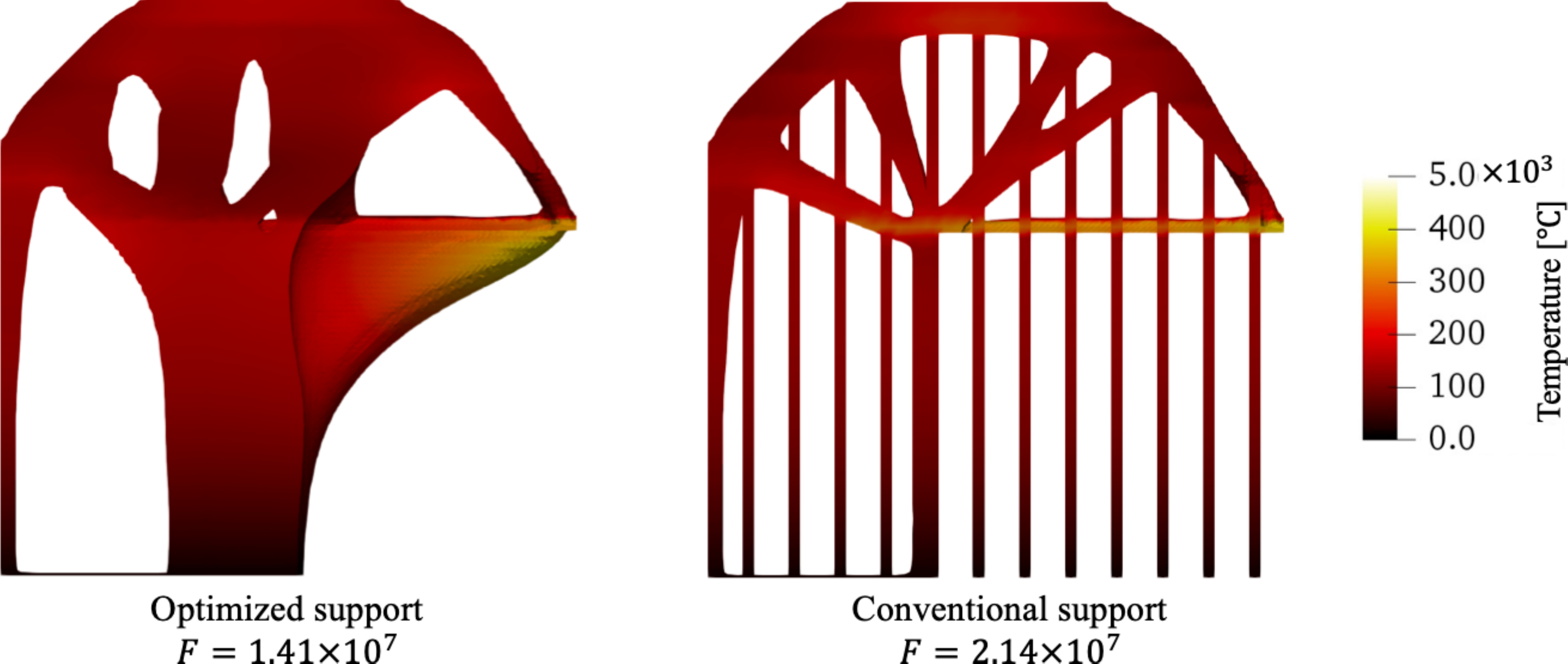}}
		\caption{Temperature field of each laser irradiation domain $\mathit{\Omega_{L}}$ in the cooling process $ t_{c}=$1 s: (a)cantilever model; (b)MBB beam model; (c)L-bracket model.}
		\label{fig:obj}
	\end{center}
\end{figure}
Figure \ref{fig:tc1} shows the temperature field in the cooling process $ t_{c}=$1 s when building the overhang region.
Each optimized support demonstrates that the maximam temperature difference is smaller than in conventional ones.
Furthermore, it is observed that the maximum temperature difference in the laser irradiation domain has also been reduced.
Figure \ref{fig:obj} displays the temperature fields of each laser irradiation domain $\mathit{\Omega_{L}}$ in the cooling process $ t_{c}=$1 s and the objective function.
It is observed that each optimized support has a more uniform temperature distribution in each laser irradiation domain than that of the conventional support, resulting in a smaller objective function.
In other words, the proposed methodology is promising for reducing thermal distortion, avoiding microstructure inhomogeneity, and degrading surface quality compared to conventional support structures.
The above results demonstrate the effectiveness of the proposed methodology for maximizing the heat dissipation in the LPBF process.
However, in the L-bracket model, the support structure has a non-uniform temperature distribution because of the overhang region.
In some cases, the support structure also needs to be included in the objective function to improve its own heat dissipation.
\section{Conclusion}\label{sec:6}
In this study, we proposed a topology optimization method for a support structure that maximizes the heat dissipation in the LPBF process.
The main contributions of this study can be summarized as follows:
\begin{enumerate}
	\item An algorithm that simulates the LPBF building process was constructed based on the transient heat conduction problem with volume heat flux.
	Through the numerical example, it was shown that the difference in heat dissipation in the laser irradiation domain appeared during the cooling process, and the overhang region had poor heat dissipation.
	\item An objective function for the support structure that maximizes heat dissipation of the part was proposed, then the optimization problem was formulated.
	The sensitivity of the objective function was derived based on the adjoint variable method and incorporated into the level-set-based topology optimization, where the level-set function was updated using the time evolutionary reaction-diffusion equation.
	In the numerical implementation, an optimization algorithm using FEM was constructed.
	\item 2D and 3D design examples were provided.
	In all models, an optimal configuration was obtained in which the support structure was added to the overhang region to improve heat dissipation.
	The improved heat dissipation in the optimal configuration was confirmed by the LPBF analytical model, demonstrating the validity and effectiveness of the proposed method.
\end{enumerate}

In future work, we aim to construct multi-objective optimization method that can also consider the part performance (e.g., compliance minimization) and optimization method that can consider the entire metal AM process including overhang limitation.
\section{Acknowledgments}
This work was supported by a JSPS grant for Scientific Research (C) JP21K03826.
\section*{References}
\bibliography{mybibfile}
\end{document}